\documentclass[12pt]{article}
\usepackage{amsmath,amsfonts}
\usepackage{graphicx}
\usepackage{enumerate}
\usepackage{soul}
\usepackage{url} 
\usepackage{caption}
\usepackage{subcaption}
\usepackage{soul}
\usepackage{color}
\usepackage{natbib}
\newcommand{\blind}{0}
\makeatletter
\newcommand*\bigcdot{\mathpalette\bigcdot@{.5}}
\newcommand*\bigcdot@[2]{\mathbin{\vcenter{\hbox{\scalebox{#2}{$\m@th#1\bullet$}}}}}
\makeatother
\definecolor{cardinal}{rgb}{0.77, 0.12, 0.23}
\newcommand{\reva}[1]{{\color{black} #1}} 

\begin{document}

\def\spacingset#1{\renewcommand{\baselinestretch}%
{#1}\small\normalsize} \spacingset{1}

\if0\blind
{
  \title{\bf Mathematical programming tools for randomization purposes in small two-arm clinical trials: A case study with real data}
  \author{Alan R. Vazquez \thanks{ORCID: \texttt{0000-0002-3658-0911}} \hspace{.2cm}\\
  \texttt{alanrvazquez@tec.mx} \hspace{.2cm}\\
    School of Engineering and Sciences, Tecnologico de Monterrey, Mexico\\
    and \\
    Weng-Kee Wong \thanks{ORCID: \texttt{0000-0001-5568-3054}} 
    \hspace{.2cm}\\
  \texttt{wkwong@ucla.edu} \hspace{.2cm}\\
    Department of Biostatistics, University of California, Los Angeles, U.S.A.}
  \maketitle
}\fi  

\if1\blind
{
  \bigskip
  \bigskip
  \bigskip
  \begin{center}
    {\LARGE\bf Mathematical programming tools for randomization purposes in small two-arm clinical trials: A case study with real data}
\end{center}
  \medskip
} \fi

\bigskip
\begin{abstract}
Modern randomization methods in clinical trials are invariably adaptive, meaning that the assignment of the next subject to a treatment group uses the accumulated information in the trial. Some of the recent adaptive randomization methods use mathematical programming to construct attractive clinical trials that balance the group features, such as their sizes and covariate distributions of their subjects. We review some of these methods and compare their performance with common covariate-adaptive randomization methods for small clinical trials. We introduce an energy distance measure that compares the discrepancy between the two groups using the joint distribution of the subjects' covariates. This metric is more appealing than evaluating the discrepancy between the groups using their marginal covariate distributions. Using numerical experiments, we demonstrate the advantages of the mathematical programming methods under the new measure. In the supplementary material, we provide R codes to reproduce our study results and facilitate comparisons of different randomization procedures.
\end{abstract}

\noindent%

\spacingset{1.45} 

\pagebreak
\section{Introduction}
\label{sec:intro}

The statistics literature is replete with randomization methods for clinical trials. Their main goal is to ensure that subjects are as comparable as possible between the treatment groups under study, so that inferences about the treatment effects are not biased by differences between the groups. In recent decades, adaptive randomization methods \citep{Rosenberger2016} have been increasingly used to achieve greater statistical efficiency and reduce costs for conducting a clinical trial. These methods assign subjects to treatment groups as they enroll in the trial using the accumulated data at hand. 

Adaptive randomization methods are classified into three groups: covariate-adaptive, response-adaptive, and covariate-adjusted response-adaptive (CARA) methods \citep{Rosenberger2016}. Covariate-adaptive methods balance the distributions of the covariates (prognostic factors) of the subjects across the treatment groups. Response-adaptive methods use the responses from subjects who have received treatment to assign future subjects to treatment groups that appear to be increasingly effective. CARA methods combine the ideas of covariate- and response-adaptive methods to generate the treatment groups of a clinical trial. We refer to \cite{Rosenberger2008HandlingTrials} and \cite{Rosenberger2012AdaptiveTrials} for complete reviews of these methods.  

In this article, we study covariate-adaptive randomization methods to balance continuous covariates in small clinical trials with two treatment groups (i.e., two-arm clinical trials). These trials arise in practice due to budget limitations or because the disease under study is rare. They also arise due to ethical and physiological considerations that may be involved in specific cohorts of subjects, such as infants \citep{Cornu2013ExperimentalChoice} or minority groups \citep{Balcazar2009AModel}. \cite{small}, \cite{reproductive} and \cite{generalized} discussed challenging issues of the design and analysis of small clinical trials. They are practically relevant but their challenges are much less addressed in the literature.

A novelty of our study is that we use mathematical programming \citep[MP;][]{bradley1977applied} to design clinical trials. MP is a systematic approach to optimizing an objective function in terms of decision variables that must satisfy a set of constraints. For this reason, it has been used frequently in engineering, operations research, and recently in the construction of various optimal designs for linear and nonlinear models \citep{duarteinfinite,duarteisi,duartemult,duarteexact,duarteadaptive,Vazquez2023ConstructingAlgorithmsb} and fractional factorial designs \citep{Bulutoglu2008ClassificationProgramming,Gromping2019AnDesigns}, as well as for analyzing data from screening experiments \citep{Phoa2009AnalysisSelector,Vazquez2021AExperiments}.  

In the MP literature, there are \reva{three adaptive} randomization methods introduced by \reva{\cite{Williamson2017ADiseases}}, \cite{Bertsimas2019Covariate-adaptiveTrials}, and \cite{Bhat2020Near-optimalTesting}. \reva{The method of \cite{Williamson2017ADiseases} is response-adaptive and assigns subjects to treatment groups using a pre-computed assignment sequence, which maximizes the expected number of correct assignments throughout the clinical trial.} The method of \cite{Bertsimas2019Covariate-adaptiveTrials} is covariate-adaptive and balances continuous covariates by matching the means and variances of their marginal distributions between groups. The method of \cite{Bhat2020Near-optimalTesting} is also covariate-adaptive and balances continuous covariates by minimizing the prediction variance of a linear regression model containing the treatment and covariate effects. An attractive feature of these methods is that they use the cumulative information on the covariate values of the subjects  already in the trial to randomize the next patient.

\reva{In this article, we compare the MP method of \cite{Bertsimas2019Covariate-adaptiveTrials} with covariate-adaptive randomization methods that fall within the minimization framework (Section~\ref{sec:review}). We chose the method of \cite{Bertsimas2019Covariate-adaptiveTrials}  over  the method of \cite{Bhat2020Near-optimalTesting} because the former is model-free. In contrast, the method of \cite{Bhat2020Near-optimalTesting}  assumes a linear model a priori, which may be incorrect, and consequently result in a suboptimal allocation scheme. To our knowledge, our comparison between minimization methods and \cite{Bertsimas2019Covariate-adaptiveTrials}'s method is new.}

We compare the methods using \reva{two} real two-arm clinical trials with up to 20 subjects and two or three continuous covariates. The performance of covariate-adaptive methods for such small trials is not addressed in the literature and so, our study fills that gap. Additionally, we propose a new measure of imbalance between two groups called the energy distance \citep{Szekely2013EnergyDistances}, which compares the joint distribution of the covariates of the subjects. Our measure is more general than the current ones that aim to balance the marginal distributions of each covariate across groups. \reva{We conclude that MP methods are more attractive than standard covariate-adaptive methods for balancing the covariate distributions and group sizes in the small two-arm clinical trials we considered.} We also show that all methods fail to balance at least one covariate in each trial, calling for method improvements for these cases. To facilitate the evaluation of the methods, we provide R codes to produce the results in the paper and may also be used to evaluate other methods. 

The remainder of the article is organized as follows. In Section~\ref{sec:review}, we review representatives of covariate-adaptive methods to balance continuous covariate distributions in two-arm clinical trials. In Section~\ref{sec:math}, we present the method of \cite{Bertsimas2019Covariate-adaptiveTrials} and, in Section~\ref{sec:examples}, we compare it with those representatives using two real clinical trials. In Section~\ref{sec:conc}, we conclude with remarks and future work directions to further investigate the utility of MP as a competitive way of randomization in clinical trials.


\section{Minimization methods}
\label{sec:review}

Covariate-adaptive randomization methods are categorized into three groups: Stratified-block randomization, dynamic hierarchical randomization, and minimization \citep{Lin2015TheTrials}. Stratified-block randomization methods \citep{Kernan1999StratifiedTrials} are two-stage procedures in which subjects who enter the clinical trial are first assigned to a stratum using their covariates' values. Within each stratum, subjects are assigned to a treatment according to a randomization procedure, such as permuted-block randomization \citep{Rosenberger2016}. Dynamic hierarchical randomization \citep{Leung1991DynamicTrials} balances the subjects' covariates following a user-specified order of importance. That is, a subject is assigned to a treatment group to balance the most important covariates first and then the least important ones. Minimization methods \citep{Taves1974Minimization:Groups} sequentially assign subjects to treatments to balance all covariates simultaneously. Specifically, they assign a subject to the treatment group that would achieve the smallest degree of imbalance across all covariates of the subjects in the trial.

\reva{We focus on minimization methods for clinical trials with two treatments and continuous covariates. These methods can be further classified into two types depending on whether a linear regression model is assumed to link the mean response to the treatments and the subjects' covariates. A method that makes such an assumption is introduced by \cite{Atkinson1982OptimumFactors,Atkinson1999OptimumInformation,Atkinson2002TheInformation,Atkinson2003TheDesigns}. This method seeks the allocation of subjects to groups that minimizes the variance of the estimated treatment effect in the linear regression model. Another method of this type is \cite{Bhat2020Near-optimalTesting} discussed previously. To limit the scope of this article, we do not include these methods in our study. Instead, we concentrate on minimization methods that do not assume a linear regression model. In particular, we study the methods of \cite{Pocock1975SequentialTrial}, \cite{Nishi2003AnGroups} and \cite{Ma2013BalancingDensities}, which are important model-free minimization methods.}

We first discuss the general minimization framework. Next, we present the discrepancy measures used by \cite{Pocock1975SequentialTrial}, \cite{Nishi2003AnGroups} and \cite{Ma2013BalancingDensities}, and then the allocation rule for new subjects. 

\subsection{The general framework} \label{sec:minimization_framework}

Minimization methods balance the marginal distributions of the covariates in the groups by minimizing a discrepancy measure. To sequentially allocate a total of $N$ subjects with $p$ covariates,  minimization methods undertake the steps below.
\begin{itemize}
    \item[Step 1.] Recruit $n_0 < N$ initial subjects and allocate them into \reva{one or two groups}.
    \item[Step 2.] For each new subject to be included in the trial, do the following:
    \begin{itemize}
        \item[2a.] Construct two potential sets of groups by hypothetically assigning the new subject to groups one and two. In one set, group one contains the subject, while in the other set, group two contains that subject. Using a discrepancy measure, compare the two sets of groups in terms of the resulting marginal distributions of the covariates.
        \item[2b.] Using the value of the discrepancy measure, determine the allocation of the subject. 
        \item[2c.] Update the treatment groups and re-start Step 2.
    \end{itemize}
    \item[Step 3.] Terminate if $N$ subjects have been recruited in the trial.
\end{itemize}

\reva{In Step 1, the value of $n_0$ can be as small as one for some minimization methods, such as \cite{Pocock1975SequentialTrial}. In this case, we randomly assign the initial subject to a group. More elaborate minimization methods, such as \cite{Nishi2003AnGroups} and \cite{Ma2013BalancingDensities}, need larger values of $n_0$ and an initial allocation to both groups. This may be achieved using random allocation or permuted-block randomization \citep{Rosenberger2016}, which ensures that each group has the same number of subjects. In this way, these methods equipoise the two groups at the beginning of the sequential allocation process in Step 2.} 

\reva{In small clinical trials, the value of $n_0$ should not be too large because the initial allocation method in Step 1 will overtake the covariate-adaptive allocation in Step 2, limiting the gain in balancing the covariates. Finding the best value of $n_0$ is a practically relevant but challenging problem that we do not address here. Instead, we follow \cite{Ma2013BalancingDensities} and use permuted-block randomization with blocks of size $n_{0}/2$ and $n_0 = 8$ for all minimization methods discussed here.} 

\subsection{Discrepancy measures}

A discrepancy measure allows us to determine the assignment of the $t$-th subject ($n_0 < t \leq N$) given that the previous $t-1$ subjects have already been assigned to the treatment groups. In what follows, we denote the treatment group as $k \in \{1,2\}$ and the $p$ subjects' covariates as $W_j$, $j =1, \ldots, p$. The sizes of groups one and two before allocating the $t$-th subject are $n_1$ and $n_2$, respectively.

\subsubsection{Difference in covariate groups}

The discrepancy measure of \cite{Pocock1975SequentialTrial} uses the difference between the number of subjects with a specific covariate value in the two groups. To use this measure, we first transform each continuous covariate $W_j$ into a discrete variable, denoted by $Z_j$, with a user-specified number of categories. For simplicity, we assume that all transformed covariates $Z_j$ have $c$ categories, which can be created using the quantiles of the distribution of $W_j$ at probabilities $1/c, 2/c, \ldots, (c-1)/c$.

Let $n_{jlk}$ be the number of subjects in group $k$ whose covariate $Z_j$ is at level $l \in \{1, \ldots, c\}$, before allocating the $t$-th subject. We denote the categories of the covariates of the $t$-th subject as $r_{1}, r_{2}, \ldots, r_{p}$, and let $n_{j r_{j} k}$ be the number of subjects in group $k$ with $Z_j$ in category $r_{j}$, before allocating the subject. \cite{Pocock1975SequentialTrial} calculate the individual discrepancy between the two groups for the $j$-th covariate as
\begin{equation*} \label{eq:PS}
    \Delta d^{PS}_{j} = |(n_{j r_j 1} + 1) - n_{j, r_j 2}| - |n_{j r_j 1} - (n_{j, r_j 2} + 1)|.
\end{equation*}
\noindent A value of $\Delta d^{PS}_{j}$ smaller than zero means that the allocation of the $t$-th subject to group one is better than that to group two. In this case, the difference between the number of subjects in category $r_j$ in the groups is smaller than the difference that would be obtained if the subject were assigned to group two. If $\Delta d^{PS}_{j}$ is larger than zero, assigning the $t$-th subject to group two is preferred. This is because this assignment reduces the gap between the number of subjects with covariate $Z_j$ in category $r_j$ in the groups, compared to assigning the subject to group one. Specifically, the discrepancy measure of \cite{Pocock1975SequentialTrial} computes the overall imbalance among all covariates using
\begin{equation}\label{eq:disc_ps}
    D_{PS} = \sum_{j=1}^{p} \Delta d^{PS}_{j}.
\end{equation}

\subsubsection{Difference in means and variances}

\cite{Nishi2003AnGroups} compare the mean and standard deviation of the marginal distributions of the covariates in the two groups. Before allocating the $t$-th subject, we consider $\bar{W}_{jk}$ and $S_{jk}$ as the mean and standard deviation, respectively, of the $j$-th covariate in group $k$. For this covariate, we define the grand mean, $\bar{W}_{j \bigcdot}$, and grand standard deviation, $S_{j \bigcdot}$, across the current groups as
\begin{equation*}
    \bar{W}_{j \bigcdot} = \frac{n_1 \bar{W}_{j1} + n_2\bar{W}_{j2} }{n_1 + n_2} \text{ and } S_{j \bigcdot} = \left(\frac{ (n_1 - 1) S^2_{j1} + (n_2 - 1) S^2_{j2} }{n_1 + n_2 - 2} \right)^{1/2}.
\end{equation*}

To assess the benefit of adding the $t$-th subject to a group, we calculate the hypothetical mean and standard deviation for the two groups including the subject. More specifically, we define $\bar{W}^{(+)}_{j1}$ and $\bar{W}^{(+)}_{j2}$ as the group means of the $j$-th covariate given that the $t$-th subject is included in groups one and two, respectively. The group standard deviations, $S^{(+)}_{j1}$ and $S^{(+)}_{j2}$, of the $j$-th covariate are defined similarly. If group one includes the subject, the grand mean and grand standard deviation across the groups would be updated with 
\begin{equation*}
    \bar{W}^{(1)}_{j \bigcdot} = \frac{(n_1+1) \bar{W}^{(+)}_{j1} + n_2\bar{W}_{j2} }{n_1 + n_2 + 1} \text{ and } S^{(1)}_{j \bigcdot} = \left(\frac{ n_1 S^{(+)^2}_{j1} + (n_2 - 1) S^2_{j2} }{n_1 + n_2 - 1} \right)^{1/2}.
\end{equation*}
\noindent If group two includes the subject, the updated mean and standard deviations are defined similarly and denoted as $\bar{W}^{(2)}_{j \bigcdot}$ and $S^{(2)}_{j \bigcdot}$, respectively. That is, the coefficients of $\bar{W}_{j2}$ and $S^2_{j2}$ in $\bar{W}^{(2)}_{j \bigcdot}$ and $S^{(2)}_{j \bigcdot}$ carry the extra one. 

\cite{Nishi2003AnGroups} measure the goodness of allocation of the $t$-th subject to group $k$ in terms of the $j$-th covariate using the function: 
\begin{equation}\label{eq:ind_imbalance_means}
    d_j(k) = |\bar{W}^{(+)}_{jk} - \bar{W}^{(k)}_{j \bigcdot}| - |\bar{W}_{jk} - \bar{W}_{j \bigcdot}| + |S^{(+)}_{jk} - S^{(k)}_{j \bigcdot}| - |S_{jk} - S_{j \bigcdot}|.
\end{equation}

The first two elements on the right-hand side of \eqref{eq:ind_imbalance_means} are the absolute distance between the new mean of group $k$ and the new grand mean if subject $t$ is in group $k$, and the absolute distance between the current mean of group $k$ and the current grand mean. A small difference between these elements is preferred because this indicates that the mean of group $k$ with the new subject is closer to the grand mean. The last two elements of \eqref{eq:ind_imbalance_means} are the absolute distance between the new group-specific and grand standard deviations if subject $t$ is in group $k$, and the absolute distance between the current group-specific and grand standard deviations. Ideally, the difference between these elements is small too, as it implies that the standard deviation of group $k$ with the new subject is close to the grand standard deviation.

We define the individual discrepancy in the $j$-th covariate as $\Delta d^{NT}_j = d_j(1) - d_j(2)$. A positive value of $\Delta d^{NT}_j$ means that assigning the $t$-th subject to group two reduces the gap between the mean and standard deviations of the distributions of covariate $j$ in the two groups, compared to assigning the subject to the other group. A negative value of $\Delta d_j$ means that the assignment of the $t$-th subject to group one results in the smallest gap.

The overall discrepancy measure of \cite{Nishi2003AnGroups} is
\begin{equation}\label{eq:disc_nt}
    D_{NT} = \sum_{j=1}^{p} \Delta d^{NT}_j + \frac{n_1 - n_2}{n_{1} + n_{2}},
\end{equation}
\noindent where the last term promotes subject assignments that create equal group sizes. So, the smaller the value of this term, the closer $n_1$ is to $n_2$.

\subsubsection{Difference in distributions}

The discrepancy measure of \cite{Ma2013BalancingDensities} compares the marginal distributions of the covariates in the two groups using a Kernel density estimate \citep{Scott2015MultivariateEdition}. It is assumed that the covariates of the subjects in the trial are standardized to have a mean of zero and a variance of one. The estimated density function for covariate $W_j$ in group $k$ is given by
\begin{equation*}
    \hat{f}_{jk}(w) = \frac{1}{n_k h(n_k)} \sum_{i=1}^{n_k} K\left(\frac{w - w'_{ijk}}{h(n_k)} \right)    
\end{equation*}
\noindent where $K(u)$ is the Kernel function, $h(v)$ the bandwidth function, and $w'_{ijk}$ the $j$-th covariate value of the $i$-th subject in group $k$ in the standardized scale. Note that the density function is estimated using the covariate values of the $t-1$ subjects already in the trial. \cite{Ma2013BalancingDensities} use the bandwidth function $h(v) = v^{-0.2}$ and the normal Kernel $ K(u) = (2 \pi)^{-1/2} \exp(-u^2/2)$.

For the $t$-th subject, the following function defines an individual discrepancy in $W_j$: \begin{equation}\label{eq:ind_imbalance_kernel}
    \Delta d^{MH}_j(w'_{tj}) = \frac{n_1}{n} \hat{f}_{j1}(w'_{tj}) - \frac{n_2}{n} \hat{f}_{j2}(w'_{tj}),
\end{equation}
\noindent where $w'_{tj}$ is subject $t$'s $j$-th covariate value on the standardized scale. \eqref{eq:ind_imbalance_kernel} can then be interpreted as a weighted difference between the probability that the $j$-th covariate value of the $t$-th subject comes from treatment one or two, where the weights are the proportions of the groups. For example, if the groups are balanced and $\Delta d^{MH}_j(w'_{tj}) >0$, the $j$-th covariate value of the new subject is more likely to follow the covariate distribution of group one, and therefore its allocation to group two is preferred to balance the distributions. If $\Delta d^{MH}_j(w'_{tj}) <0$, the $j$-th covariate value of the new subject is more likely to follow the covariate distribution of group two and so, its allocation to group one is preferred.

The overall discrepancy measure of \cite{Ma2013BalancingDensities} for all covariates is 
\begin{equation} \label{eq:disc_ma}
    D_{MH} = \sum_{j=1}^{p} \Delta d^{MH}_j(w'_{tj}).
\end{equation}

\subsection{Decision rule}

In the minimization framework, the final allocation of the new subject is typically determined using the biased-coin strategy of \cite{Efron1971ForcingBalanced}. To this end, we first set a user-specified probability, $P_0$, ranging from 1/2 to one. Next, we assign the new subject to group one or two with probability $P$ and $1-P$, respectively, where $P$ depends on $P_0$ and $D$, the discrepancy measure value of the subject. Specifically, we have that
\begin{equation} \label{eq:balanced_coin}
    P = \begin{cases}
    P_0 & \text{ if } D < 0, \\
    1 - P_0 &  \text{ if } D > 0, \\
    1/2 & \text{ otherwise},
    \end{cases}
\end{equation}
\noindent where we compute $D$ using either \eqref{eq:disc_ps}, \eqref{eq:disc_nt}, \eqref{eq:disc_ma}, or \eqref{eq:bertsimas_discrepancy} for the method of \cite{Bertsimas2019Covariate-adaptiveTrials} discussed later.

\reva{The biased-coin strategy allows a minimization method to have randomness in allocating subjects to groups and maintain approximately balanced groups in terms of the covariates \citep{Berger2021ATrials}. Allocation randomness reduces the possibility that the user, say, a physician, knows the allocation rule in advance and interferes with it. For example, the user may ignore the rule and assign the subject to the treatment that, in the user's opinion, would benefit the subject. In this case, the user commits \textit{selection bias} \citep{Rosenberger2016}, which may invalidate the statistical conclusions of the trial.} 

\reva{If $P_0 = 1/2$ in \eqref{eq:balanced_coin}, the assignment of the subject to a group is completely random, thus avoiding selection bias. However, this value can cause a severe imbalance in the covariates between the groups at the end of the recruitment. A value of $P_0$ equal to one results in groups with balance in their covariates, but a high risk of selection bias. This is because the assignment is deterministic when the discrepancy measure value of the subject differs from zero. A value of $P_0$ between 1/2 and one is thus ideal as it gives the trade-off between covariate balance and allocation randomness. The value should be such that it biases the assignment towards the best group according to the discrepancy measure (to promote covariate balance) and provides a good degree of allocation randomness (to mitigate selection bias). Based on the simulation study of \cite{Toorawa2009UseRandomization} on minimization methods, we use $P_0 = 0.8$ unless otherwise stated.}  

\section{A minimization method rooted in mathematical programming}
\label{sec:math}

The covariate-adaptive method of \cite{Bertsimas2019Covariate-adaptiveTrials} consists of solving a subclass of MP problems called mixed-integer programming (MIP) problems \citep{Bertsimas2005OptimizationIntegers}. In a MIP problem, the goal is to determine the values of a set of discrete and continuous decision variables to optimize a linear objective function while satisfying a set of constraints. A MIP problem formulation has the form:  
\begin{subequations}
\begin{align}
&\max_{\mathbf{x}}   \;\;\; \mathbf{c}^{T} \mathbf{x} \label{eq:intro_objective}  \\ 
\text{ subject to}  \nonumber \\
 & \mathbf{G} \mathbf{x} = \mathbf{b}, \label{eq:intro_equal}  \\  
 & \mathbf{H} \mathbf{x} \leq \mathbf{d} ,  \label{eq:intro_ineq}  \\ 
		& x_i \in \mathbb{Z}, \;\; \forall i \in \mathcal{J},  \nonumber \label{eq:intro_integer} 
\end{align}
\end{subequations}
\noindent where $\mathbf{x} = (x_1, x_2, \ldots, x_n)^{T}$ is an $n \times 1$ vector of decision variables, $\mathbf{c}$ is an $n \times 1$ constant vector, $\mathbf{G}$ is an $m_1 \times n$ constant  matrix, $\mathbf{H}$ is an $m_2 \times n$ constant matrix, $\mathbf{b}$ is an $m_1 \times 1$ constant  vector, $\mathbf{d}$ is an $m_2 \times 1$ constant vector, and $\mathcal{J}$ is a nonempty set of indices. 

\reva{The above formulation is general and can be adapted to reflect the objective of a clinical trial. For example, in the context of non-adaptive designs for two-arm clinical trials, the entries of $\mathbf{x}$ can take the value $x_i = 1$ if the $i$-th subject is assigned to group one, and the value $-1$, otherwise. If $n$ is even, the equality constraint in \eqref{eq:intro_equal} can indicate that the group sizes are equal to $n/2$. This can be done by setting $\mathbf{b} = 0$ and $\mathbf{G} = \mathbf{1}^T_n$, where $\mathbf{1}_n$ is the $n \times 1$ vector with all entries equal to 1. If $n$ is odd, \eqref{eq:intro_ineq} can indicate that the absolute difference in the group sizes does not exceed one. We can achieve this by setting $\mathbf{d} = \mathbf{1}_2$ and $\mathbf{H}$ to the $2 \times n$ matrix $(\mathbf{1}_n; -\mathbf{1}_n)^T$. In this case, we should also set $\mathbf{G} = \mathbf{0}^T_n$ to remove \eqref{eq:intro_equal}, where $\mathbf{0}_n$ is the $n \times 1$ vector with all entries equal to 0. A goal of the MIP problem may be to balance the average of the values of a covariate between the groups. To this end, the $i$-th entry of $\mathbf{c}$ in \eqref{eq:intro_objective} is set to the covariate value of the $i$-th subject. The solution to this MIP problem is then the assignment of subjects to groups that minimizes the difference between the average covariate values between them and satisfies the constraints.}  

Commercial optimization software such as GUROBI, MOSEK, and BARON can solve several types of MIP problems using state-of-the-art optimization techniques \citep{Junger201050State-of-the-art}. Throughout the solution process, the software provide both the best feasible solution and the best bound for the objective function's optimal value obtained so far. If the objective value of the best solution is equal to the best bound, the solution is optimal.

We present the covariate-adaptive randomization method of \cite{Bertsimas2019Covariate-adaptiveTrials} and show that it falls within the minimization framework in Section~\ref{sec:minimization_framework}. This close link between MP and minimization methods is new to the literature. 

\subsection{Incorporating prior information on the covariates of future subjects} \label{sec:uncertainty}

We denote the $p \times 1$ covariate vector of the $i$-th subject as $\mathbf{w}_{i} = (w_{i1}, w_{i2}, \ldots, w_{ip})^{T}$,  where $w_{ij}$ is the $j$-th covariate value. We define the $p \times 1$ sample mean vector $\bar{\mathbf{w}}_t$ and the $p \times p$ empirical covariance matrix $\Sigma_t$ of the covariate vectors available before allocating the ($t+1$)-th subject as:
\begin{equation*}
    \bar{\mathbf{w}}_t = \frac{1}{t} \sum_{i=1}^{t}  \mathbf{w}_i \text{  and  } \Sigma_{t} = \frac{1}{t} \sum_{i=1}^{t} (\mathbf{w}_i - \bar{\mathbf{w}}_t)(\mathbf{w}_i - \bar{\mathbf{w}}_t)^{T}.
\end{equation*}

To incorporate prior information on future subjects, \cite{Bertsimas2019Covariate-adaptiveTrials} define the unknown covariate vectors for these subjects as $\tilde{\mathbf{w}}_i$ and their entries as $\tilde{w}_{ij}$, with $i = t+1, \ldots, N$ and $j = 1, \ldots, p$. These vectors are collected in the $p \times (N-t)$ covariate matrix $\tilde{\mathbf{W}} = [\tilde{\mathbf{w}}_{t+1}; \cdots; \tilde{\mathbf{w}}_N]$ that belongs to the set 
\begin{equation*}
    U_{\tilde{W}} = \left\{ \tilde{\mathbf{W}} \in \mathbb{R}^{p \times (N-t)} \; \big| \; \tilde{\mathbf{w}}_i = \bar{\mathbf{w}}_t + (\Sigma_t)^{1/2} \boldsymbol{\epsilon}_i, \; i = t+1, \ldots, N,\; \mathbf{E} \in U_E \right\},
\end{equation*}
\noindent where the $p \times 1$ vector $\boldsymbol{\epsilon}_i = (\epsilon_{i1}, \ldots, \epsilon_{ip})^{T}$ contains a random perturbation to the $j$-th covariate value of the $i$-th subject, denoted by $\epsilon_{ij}$. The $p \times (N-t)$ matrix $\mathbf{E} = [\boldsymbol{\epsilon}_{t+1}; \cdots; \boldsymbol{\epsilon}_{N}]$ that collects all perturbation vectors belongs to the uncertainty set:
\begin{equation*}
    U_E = \left\{ \mathbf{E} \in \mathbb{R}^{p \times (N-t)} \; \big| \; \| \mathbf{E} \|_2 \leq \Gamma \sqrt{(N-t)p} \right\},
\end{equation*}
\noindent where $\| \mathbf{E} \|_2$ is the Frobenious norm of $\mathbf{E}$.

The tuning parameter $\Gamma$ in $U_E$ controls how diffuse the information on the future subjects' covariate vectors is, with respect to the covariate vectors of the subjects in the trial. A small $\Gamma$ value implies that the covariate vectors of the future subjects will be close to the covariate vector of the ``average'' subject in the trial. A large $\Gamma$ value implies that future subjects' covariate vectors may heavily depart from this subject. \cite{Bertsimas2019Covariate-adaptiveTrials} suggest to choose a value of $\Gamma$ between 1/2 and four. \reva{This recommendation is supported by their numerical experiments involving real and synthetic clinical trials.}

\subsection{Problem formulation} \label{sec:formulation}

The problem formulation of \cite{Bertsimas2019Covariate-adaptiveTrials} for covariate-adaptive allocation determines the assignments of current and future subjects that minimize the absolute difference in the means and variances of the covariates between the two treatment groups. It is assumed that $N$ is even and that both groups will end up with $N/2$ subjects. In their problem formulation, the assignment of the $i$-th subject to a group is expressed using the binary decision variable $x_i$, where $i = t, \ldots, N$. The variable $x_i$ takes the value of one if and only if the subject is assigned to group one. Otherwise, it takes a value of zero, and the subject is assigned to group two. Regarding the subjects already in the trial, we also denote their group allocations using binary variables. Specifically, we use the binary variable $\hat{x}_i$ which takes a value of one or zero if and only if the subject is in group one or two, respectively, where $i = 1, \ldots, t-1$. 

Before allocating the $t$-th subject, we consider the mean of the $j$-th covariate from group $k$, denoted by $\bar{W}_{jk}$, which would be obtained at the end of the recruitment of subjects. Using the binary encoding for the group assignments, the difference in the mean of the $j$-th covariate between the  groups is
\begin{equation*}
    \bar{W}_{j1} - \bar{W}_{j2} = \frac{2}{N} \left\{ \sum_{i=1}^{t-1} w_{ij} (2\hat{x}_i - 1) + w_{tj} (2x_{t} - 1) + \sum_{i=t+1}^{N} \tilde{w}_{ij} (2x_i -1) \right\}.
\end{equation*}

\cite{Bertsimas2019Covariate-adaptiveTrials} compute an \textit{approximated} variance of the $j$-th covariate under group $k$, denoted by $V_{jk}$, which would be obtained at the end of the assignment of subjects. In this approximation, the covariate vectors are centered around the $j$-th entry of the vector $\bar{\mathbf{w}}_t$, instead of the mean vector of all subjects in the trial. In this way, we limit the amount of uncertainty in the calculations of the group variances produced by the unknown covariate vectors of future subjects. The difference between approximated variances of the $j$-th covariate under groups one and two is
\begin{align*}
    V_{j1} - V_{j2} = \frac{2}{N} & \left\{ \sum_{i=1}^{t-1} (w_{ij} - \bar{w}_{tj} )^2 (2\hat{x}_i - 1) + (w_{tj} - \bar{w}_{tj})^2 (2x_{t} - 1) \right.  \\  & \left.+ \sum_{i=t+1}^{N} (\tilde{w}_{ij} - \bar{w}_{tj})^2 (2x_i -1) \right\},
\end{align*}
\noindent where $\bar{w}_{tj}$ is the $j$-th entry of $\bar{\mathbf{w}}_t$.

The problem formulation to obtain the optimal assignment of the $t$-th subject is
\begin{subequations}
\begin{align}
\min_{\mathbf{x}, \mathbf{W}, \mathbf{V}}   &\;\;\; \sum_{j=1}^{p} W'_{j\bigcdot} + \rho V_{j\bigcdot} \label{eq:objective} \\
\text{subject to} & 	\nonumber \\
		& W'_{j\bigcdot} = | \bar{W}_{j1} - \bar{W}_{j2} |, \quad j = 1, \ldots, p, \quad \forall \tilde{\mathbf{W}} \in U_{\tilde{W}}. \label{cons:difference_means}\\
		& V_{j\bigcdot} = | V_{j1} - V_{j2} |, \quad j = 1, \ldots, p, \quad \forall \tilde{\mathbf{W}} \in U_{\tilde{W}} \label{cons:difference_std_dev}\\
             & \sum_{i=1}^{t-1} \hat{x}_i + x_t + \sum_{i=t+1}^{N} x_{i} = \frac{N}{2}, \label{cons:cardinality} \\
             & x_{i} \in \{0, 1\}, \quad i = t, \ldots, N.  \label{cons:binary}
\end{align}
\end{subequations}
\noindent This problem formulation has $N-t$ binary decision variables in $\mathbf{x} = (x_{t}, x_{t+1}, \ldots, x_{N})^{T}$, $2p$ continuous decision variables in $\mathbf{W} = (W'_{1\bigcdot}, W'_{2\bigcdot}, \ldots, W'_{p\bigcdot})^{T}$ and $\mathbf{V} = (V_{1\bigcdot}, V_{2\bigcdot}, \ldots, V_{p\bigcdot})^{T}$, $2p$ constraints involving the unknown covariate vectors $\tilde{\mathbf{w}}$ in   \eqref{cons:difference_means} and  \eqref{cons:difference_std_dev}, and a linear constraint in \eqref{cons:cardinality}. Technically, the problem formulation is a robust MIP problem \citep{Ben-Tal2003RobustProblems}, because the values of some of the constants in  \eqref{cons:difference_means} and \eqref{cons:difference_std_dev} are uncertain.

The objective function in \eqref{eq:objective} is a weighted sum of the absolute differences in the means and the approximated variances between the groups for all covariates. The weights are determined by the tuning parameter $\rho$, which controls the trade-off between the imbalance in the group means and the approximated group variances. A large value of $\rho$ emphasizes a close match between the approximated variances of the group in the covariates, while a small value emphasizes a close match between the means of the groups.   

The constraints in  \eqref{cons:difference_means} and \eqref{cons:difference_std_dev} give the absolute differences between the group means and the approximated group variances of the covariates, respectively. The constraint in \eqref{cons:cardinality} states that the size of group one must be $N/2$ at the end of the trial. It also implies that the size of group two must also be $N/2$. The constraints contained within \eqref{cons:binary} ensure that the decision variables $x_{i}$ are binary.

\subsection{Reformulation as a minimization method}

\cite{Bertsimas2019Covariate-adaptiveTrials} show that the problem formulation in \eqref{eq:objective}--\eqref{cons:binary} can be reformulated as a standard MIP problem, whose solution involves the evaluation of two feasible solutions only. More specifically, we need to evaluate the two sets of potential groups that would result if the $t$-th subject is allocated to group one or two. Here, we show that the covariate-adaptive randomization method of \cite{Bertsimas2019Covariate-adaptiveTrials} is actually a minimization method. 

To see this, we first let $\mathbf{v}_{j}$ be the $j$-th row of $(\Sigma_t)^{1/2}$. In the alternative problem formulation of \citet[][sec.  EC.2]{Bertsimas2019Covariate-adaptiveTrials}, the absolute difference between the group means for the $j$-th covariate in \eqref{cons:difference_means} reduces to
\begin{equation} \label{eq:new_diff_means}
    \frac{N}{2} W'_{j\bigcdot} =  \left| \sum_{i=1}^{t-1} (w_{ij} - \bar{w}_{tj}) (2\hat{x}_i - 1) + (w_{tj} - \bar{w}_{tj}) (2x_{t} - 1)\right| + \Gamma \| \mathbf{v}_{j} \|_2 (N-t) \sqrt{p},
\end{equation}
\noindent where $x_t \in \{0, 1\}$ is the only decision variable. The uncertainty in the covariate values of future subjects is in the last term in \eqref{eq:new_diff_means}.

The expression for the absolute difference between the approximated group variances in the $j$-th covariate ($V_{j\bigcdot}$) is slightly more complicated; see \citet[][sec.  EC.2]{Bertsimas2019Covariate-adaptiveTrials}. There is one expression for the case where $p=1$, and another for the case where $p \geq 2$. We adopt the latter because our focus is on balancing several covariates simultaneously. In this case, $V_{j\bigcdot}$ is the maximum of the quantities:
\begin{align*}
     & \frac{2}{N} \sum_{i=1}^{t-1} (w_{ij} - \bar{w}_{tj} )^2 (2\hat{x}_i - 1) + (w_{tj} - \bar{w}_{tj})^2 (2x_{t} - 1) + \tilde{\Gamma} \|\mathbf{v}_{j}\|_{2}^2\; I(k - n_1 - x_{t} - 1) \\ & \text{ and  } \\ &  -\frac{2}{N} \sum_{i=1}^{t-1} (w_{ij} - \bar{w}_{tj} )^2 (2\hat{x}_i - 1) - (w_{tj} - \bar{w}_{tj})^2 (2x_{t} - 1) - \tilde{\Gamma} \|\mathbf{v}_{j}\|_{2}^2\; I(k -
    n_2 - x_{t} - 1),
\end{align*}
\noindent where $\tilde{\Gamma} = \Gamma^2 (N-t) p$, $n_1 = \sum_{i=1}^{t-1} \hat{x}_i$, $n_2 = t - 1 - \sum_{i=1}^{t-1} \hat{x}_i$, and $I(z)$ is an indicator function that equals one if $z \geq 0$ and zero otherwise. As with the difference in group means, the only decision variable in these expressions is $x_t$ and the uncertainty in the future covariate values is in their last terms.

Since the minimization of $V_{j\bigcdot}$ is straightforward, \cite{Bertsimas2019Covariate-adaptiveTrials} use the difference in the group (approximated) standard deviations instead, so as to put the two competing objectives on the same scale and facilitate an intuitive choice of the tuning parameter $\rho$. Therefore, the discrepancy measure of these authors is
\begin{equation} \label{eq:bertsimas_discrepancy}
    D_B = \sum_{j=1}^{p} W'_{j\bigcdot} + \rho \sum_{j=1}^{p} V^{1/2}_{j\bigcdot}.
\end{equation}
\noindent \reva{Using numerical experiments, \cite{Bertsimas2019Covariate-adaptiveTrials} found that $\rho = 6$ generally leads to a good balance in terms of the group means and variances.} 

The minimization method follows the steps in Section~\ref{sec:minimization_framework} with few modifications. In Step 2, for each new subject to be included in the trial, we set a value of the uncertainty parameter $\Gamma$ at random from the interval $[1/2, 4]$. The numerical experiments of \cite{Bertsimas2019Covariate-adaptiveTrials} show that choosing the value of $\Gamma$ independently at random at each time step protects against selection bias. Since the selection process of $\Gamma$ incorporates randomness in the decision process, we determine the final allocation of the subject using $P_0 = 1$ in \eqref{eq:balanced_coin}.

\section{Numerical comparisons}
\label{sec:examples}

In this section, we compare the minimization method of \cite{Bertsimas2019Covariate-adaptiveTrials} with those in Section~\ref{sec:review} using two real clinical trials. First, we introduce the metrics to assess the balance in the covariate distributions across the groups and the randomness in the treatment allocation of the methods. Next, we show the performance of the methods for each trial separately \reva{and end the section with a discussion.}

We refer to the minimization methods of \cite{Pocock1975SequentialTrial}, \cite{Nishi2003AnGroups}, \cite{Ma2013BalancingDensities} and \cite{Bertsimas2005OptimizationIntegers} as PS, NT, MH and BKW, respectively. For the PS method, we use $c = 3$ categories for the surrogate discrete covariates. The supplementary materials accompanying the article have an R implementation of all the minimization methods discussed here. 

\subsection{Evaluation metrics}

\subsubsection{Covariate balance}

We use several evaluation metrics to measure the balance in the distributions of the covariates between the two treatment groups. They include the absolute difference in the size of the groups, the absolute difference between the means of the $j$-th covariate across the groups, and the absolute difference between the standard deviations of the $j$-th covariate across the groups. Ideally, the values of these metrics are equal or close to zero. 

The differences between the individual means and standard deviations for each individual covariate are \textit{marginal} metrics, because they measure the discrepancy between the marginal distributions of the covariates in the two groups. However, they do not measure the discrepancies between the joint distributions of the covariates that generate the data. Balance in the joint covariate distribution is more general than balance in the marginal distribution of each covariate among the groups. This is because a joint covariate distribution balance implies a balance in each marginal covariate distribution, regardless of whether the covariates are dependent or independent. In contrast, a balance in each marginal covariate distribution may only achieve a balance in the joint covariate distributions of the groups when the covariates are independent. 

To measure the discrepancies in the joint distribution of the covariates among the groups, we use the energy distance \citep{Szekely2013EnergyDistances}. Let $\mathbf{W}_k$ be the $N_k \times p$ covariate matrix for the $k$-th group and $\mathbf{w}_{ik}$ be the $i$-th row of this matrix, where $N_k$ is the size of the $k$-th group at the end of the trial and $k = 1,2$. The energy distance for two independent random samples is
\begin{equation} 
    E = \frac{2}{N_1 N_2} \sum_{u=1}^{N_1} \sum_{v=1}^{N_2} \|\mathbf{w}_{u1} - \mathbf{w}_{v2} \| - \frac{1}{N_{1}^2} \sum_{u=1}^{N_1} \sum_{l=1}^{N_1} \|\mathbf{w}_{u1} - \mathbf{w}_{l1} \| - \frac{1}{N_{2}^2} \sum_{v=1}^{N_2} \sum_{m=1}^{N_2} \|\mathbf{w}_{v2} - \mathbf{w}_{m2} \|,
\end{equation}
\noindent where $\| \mathbf{u} \|$ is the $L_2$-norm of vector $\mathbf{u}$. 

\cite{Szekely2004TestingDimension} use the statistic $E$ to test the homogeneity of two multivariate joint distributions \citep{Szekely2004TestingDimension}. Using a permutation test, they reject the hypothesis of equal joint distributions when the value of $E$ is large. Therefore, in our context, low values of the energy distance indicate that the joint covariate distributions of the groups are similar. To the best of our knowledge, the energy distance is the first metric to evaluate minimization methods in terms of the balance in the joint covariate distributions between the groups they produce.

\reva{The energy distance can assess the balance in the distributions of many covariates simultaneously using a single number. This is in contrast to marginal metrics such as the absolute difference between the group means or group standard deviations of the covariates. To evaluate the quality of a clinical trial in terms of these metrics, we must compare several values simultaneously, which can be cumbersome in the presence of many covariates. Therefore, the energy distance provides the user with a simple yet effective summary of the covariate balance in a trial, compared to marginal metrics.}

\subsubsection{Allocation randomness}

We evaluate the randomness of allocation of a minimization method using the Correct Guess (CG) probability \reva{\citep{Blackwell1957DesignBias, Zhao2012QuantitativeRandomness}}. Recall that the sizes of groups one and two before allocating the $t$-th subject are $n_1$ and $n_2$, respectively, and that, in the problem formulation of \cite{Bertsimas2019Covariate-adaptiveTrials}, $x_t$ is one if the $t$-th subject is assigned to group one and it is zero otherwise. At time step $t$, the CG probability is
\begin{equation*}
\text{CG}_t = \begin{cases}
1 & \text{if } (n_{1} < n_{2} \text{ and } x_t = 1) \text{ or } (n_{1} > n_{2} \text{ and } x_t = 0) \\
1/2 & \text{if } (n_{1} = n_{2}) \\
0 & \text{if } (n_{1} > n_{2} \text{ and } x_t = 1) \text{ or } (n_{1} < n_{2} \text{ and } x_t = 0)
\end{cases}.
\end{equation*}
\noindent The value of CG$_t$ is one if the group sizes are unequal and we can determine the assignment of the $t$-th subject using the smallest group size. If the group sizes are unequal, but we cannot determine the assignment using this rule, the value of CG$_t$ is zero. If the group sizes are equal at time $t$, CG$_t$ is 1/2.  

We compute the CG probability for all treatment allocations in Step 2 of the minimization framework in Section~\ref{sec:minimization_framework}. That is, we calculate CG$_t$ for $t = n_0+1, n_0+2, \ldots, N$. We use the mean of the CG$_t$ values as an overall measure of allocation randomness \citep{Zhao2012QuantitativeRandomness}. We prefer a low mean CG probability because this implies a low mean probability of guessing the assignment of the next subject using the previous allocations. \reva{Simple randomization---in which we assign each subject to a group completely at random and independently of the previous assignments---has a mean CG probability of 1/2.}

\subsection{The pembrolizumab clinical trial} \label{sec:pembro_trial}

\cite{Sundahl2019RandomizedPresented.} studied the safety of pembrolizumab when combined with sequential or concomitant body radiotherapy in metastatic bladder cancer. The response under study was the dose-limiting toxicity measured on a specific scale. The clinical trial consisted of 18 subjects whose prognostic factors included age, sex, hemoglobin concentration, a modified proportion score of PD-L1, smoking status, among others. The subjects were enrolled into two equally sized groups. Groups one and two were administered pembrolizumab using sequential and concomitant body ratiotherapy, respectively. 

We compare the actual assignment of the subjects to the two groups of \cite{Sundahl2019RandomizedPresented.} with the minimization methods to balance the continuous covariates: age, the modified proportion score of PD-L1, and hemoglobin concentration. To evaluate the methods, we standardized the covariates to have a mean of zero and a standard deviation of one. We then applied each method 1,000 times to balance the covariates in the standardized set. 

\begin{figure}[b]
\centering
    \includegraphics[width=0.8\textwidth]{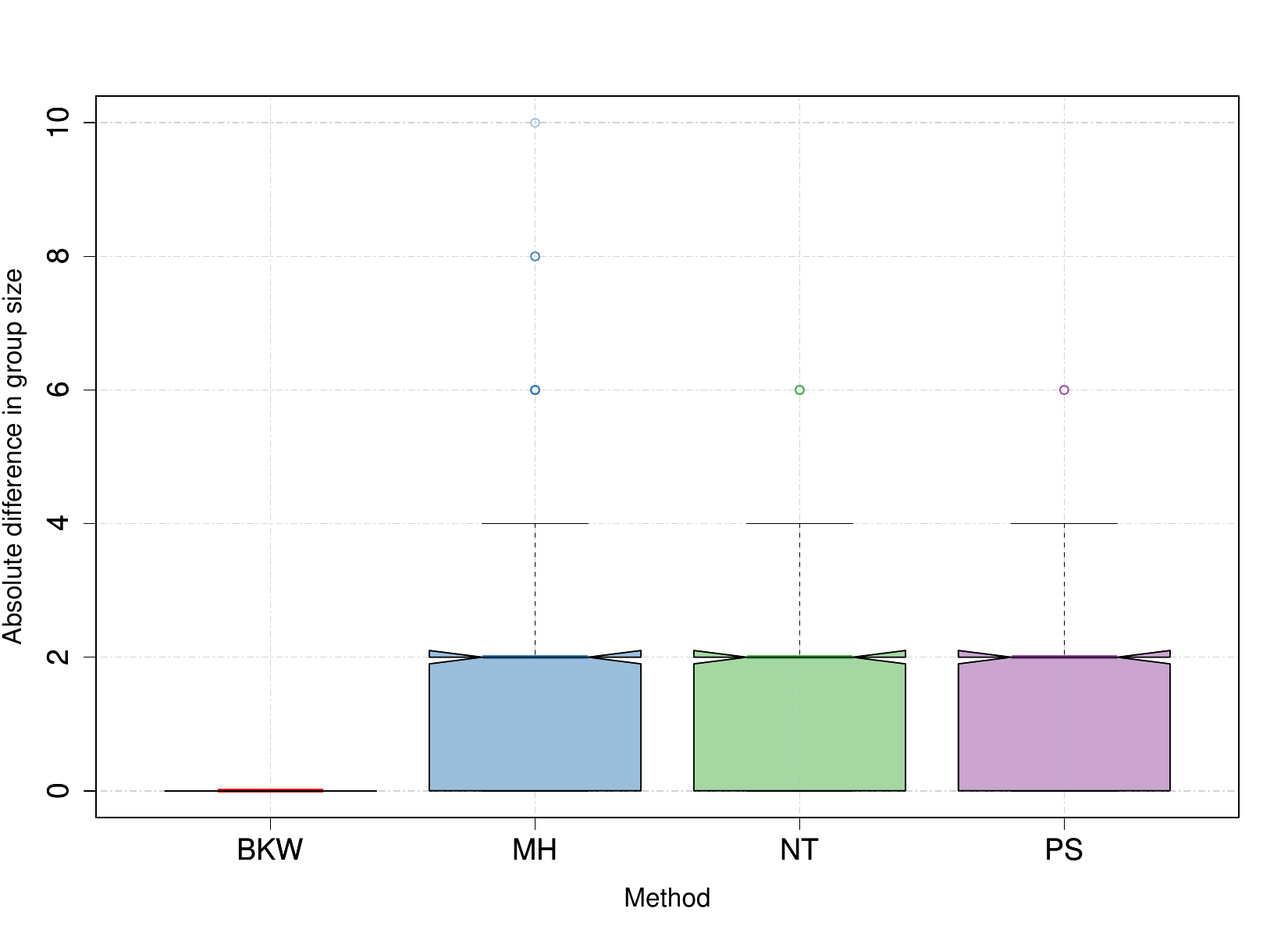}
    \caption{Boxplots of the absolute difference in group sizes obtained for the pembrolizumab trial.}
 \label{fig:pembro_group_balance}
\end{figure}

\begin{figure}[h]
     \centering
     \begin{subfigure}[b]{0.48\textwidth}
         \includegraphics[width=\textwidth]{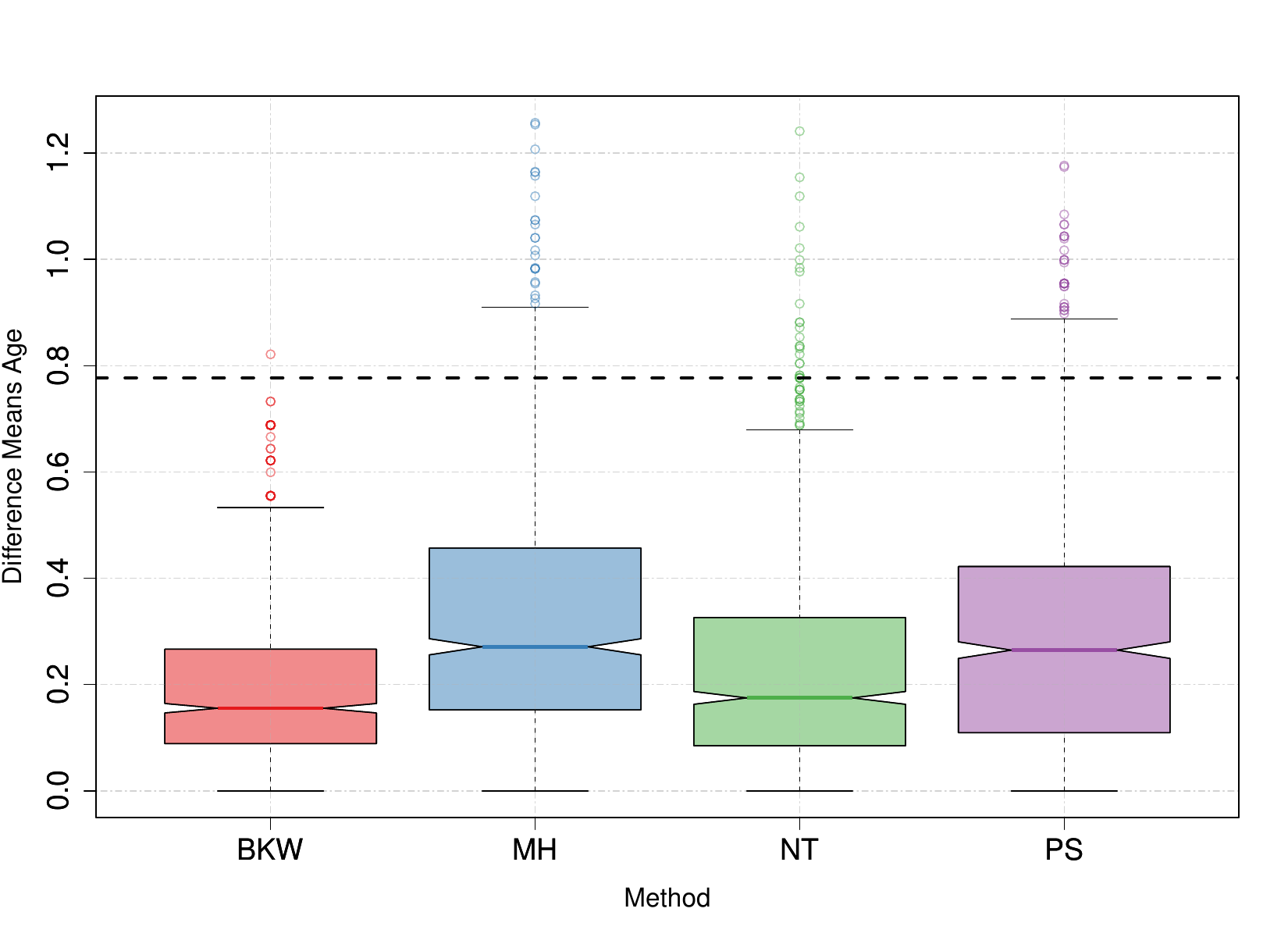}
         \caption{Absolute difference in means}
         \label{fig:pembro_age_mean}
     \end{subfigure}
     ~
     \begin{subfigure}[b]{0.48\textwidth}
         \includegraphics[width=\textwidth]{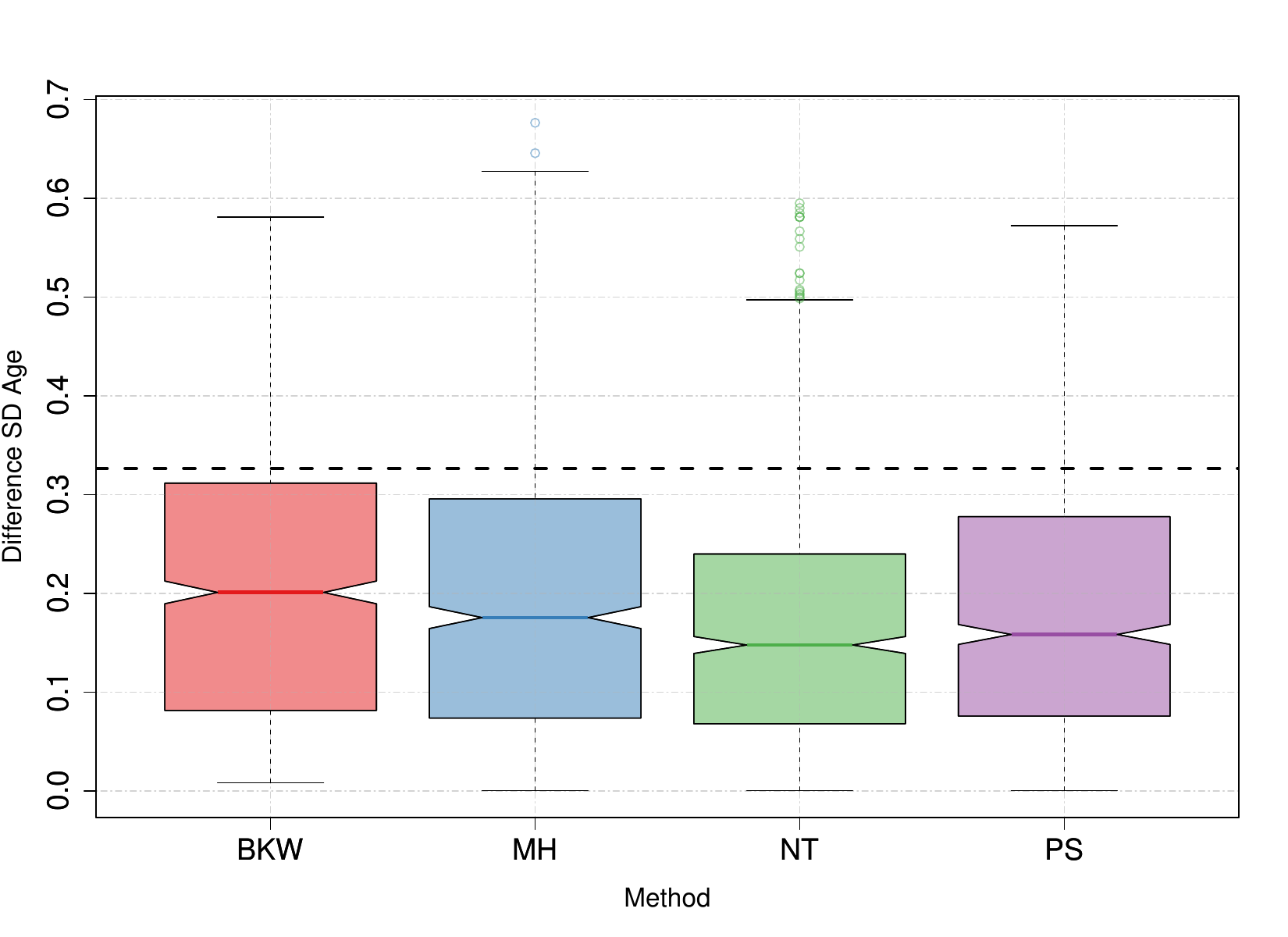}
         \caption{Absolute difference in standard deviations}
         \label{fig:pembro_age_sd}
     \end{subfigure}
        \caption{Boxplots of the absolute differences in group means and standard deviations for age in the pembrolizumab trial. The dashed line shows the actual absolute differences.}
        \label{fig:pembro_age}
\end{figure}

\begin{figure}[h]
     \centering
     \begin{subfigure}[b]{0.48\textwidth}
         \includegraphics[width=\textwidth]{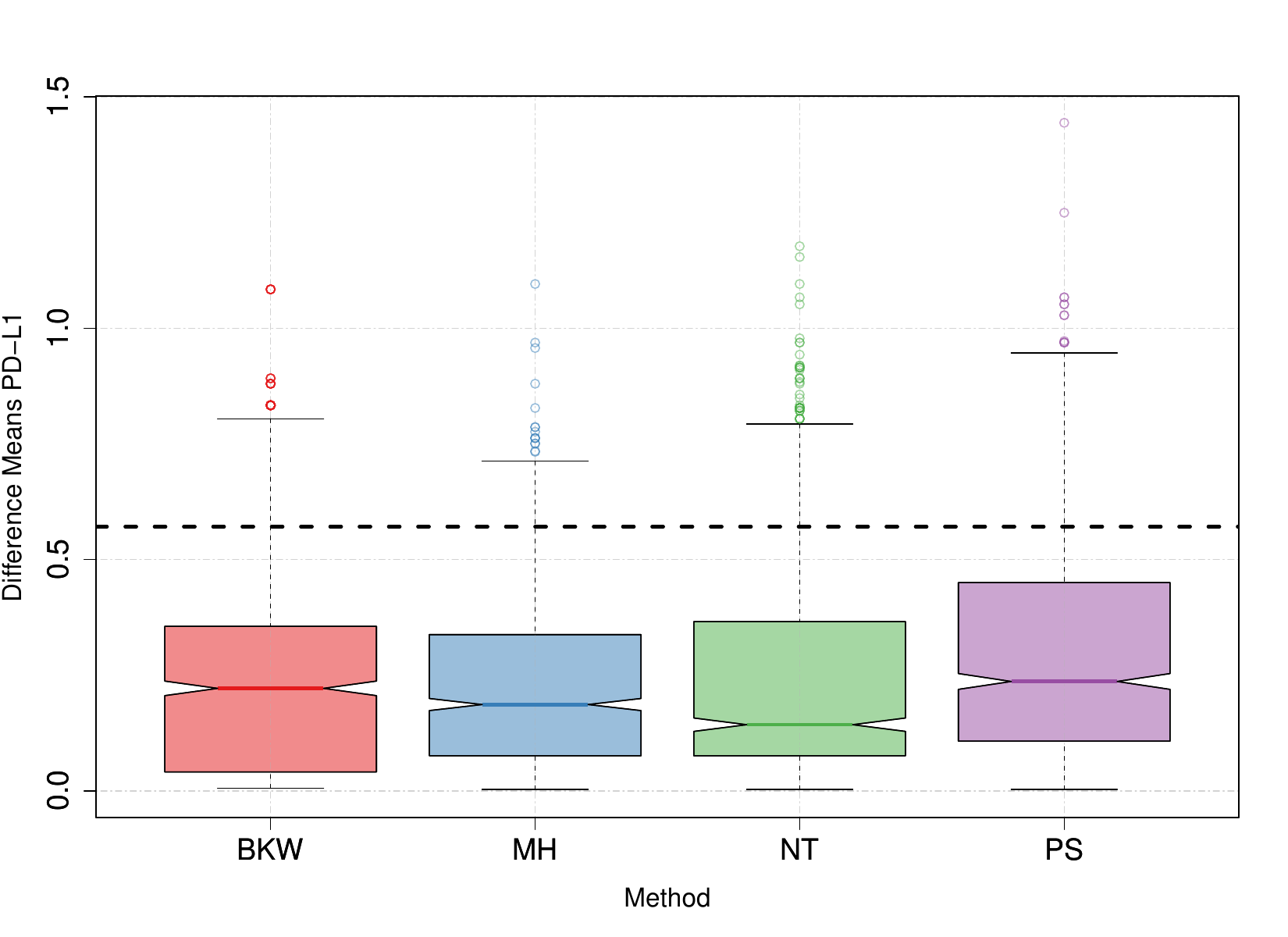}
         \caption{Absolute difference in means}
         \label{fig:pembro_PD_mean}
     \end{subfigure}
     ~
     \begin{subfigure}[b]{0.48\textwidth}
         \includegraphics[width=\textwidth]{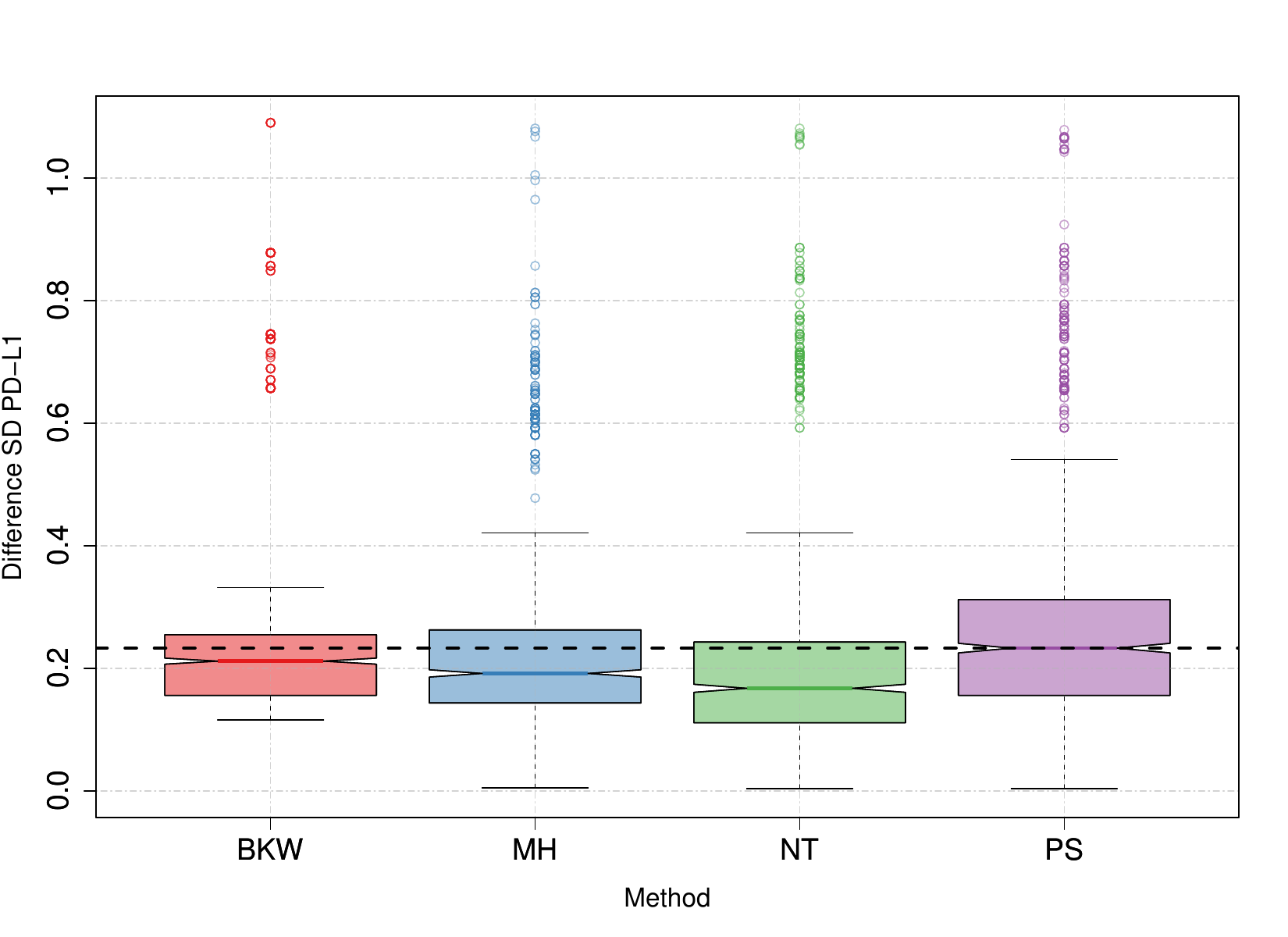}
         \caption{Absolute difference in standard deviations}
         \label{fig:pembro_PD_sd}
     \end{subfigure}
        \caption{Boxplots of the absolute differences in group means and standard deviations for PD-L1 in the pembrolizumab trial. The dashed line shows the actual absolute differences.}
        \label{fig:pembro_PD}
\end{figure}

\begin{figure}[h]
     \centering
     \begin{subfigure}[b]{0.48\textwidth}
         \includegraphics[width=\textwidth]{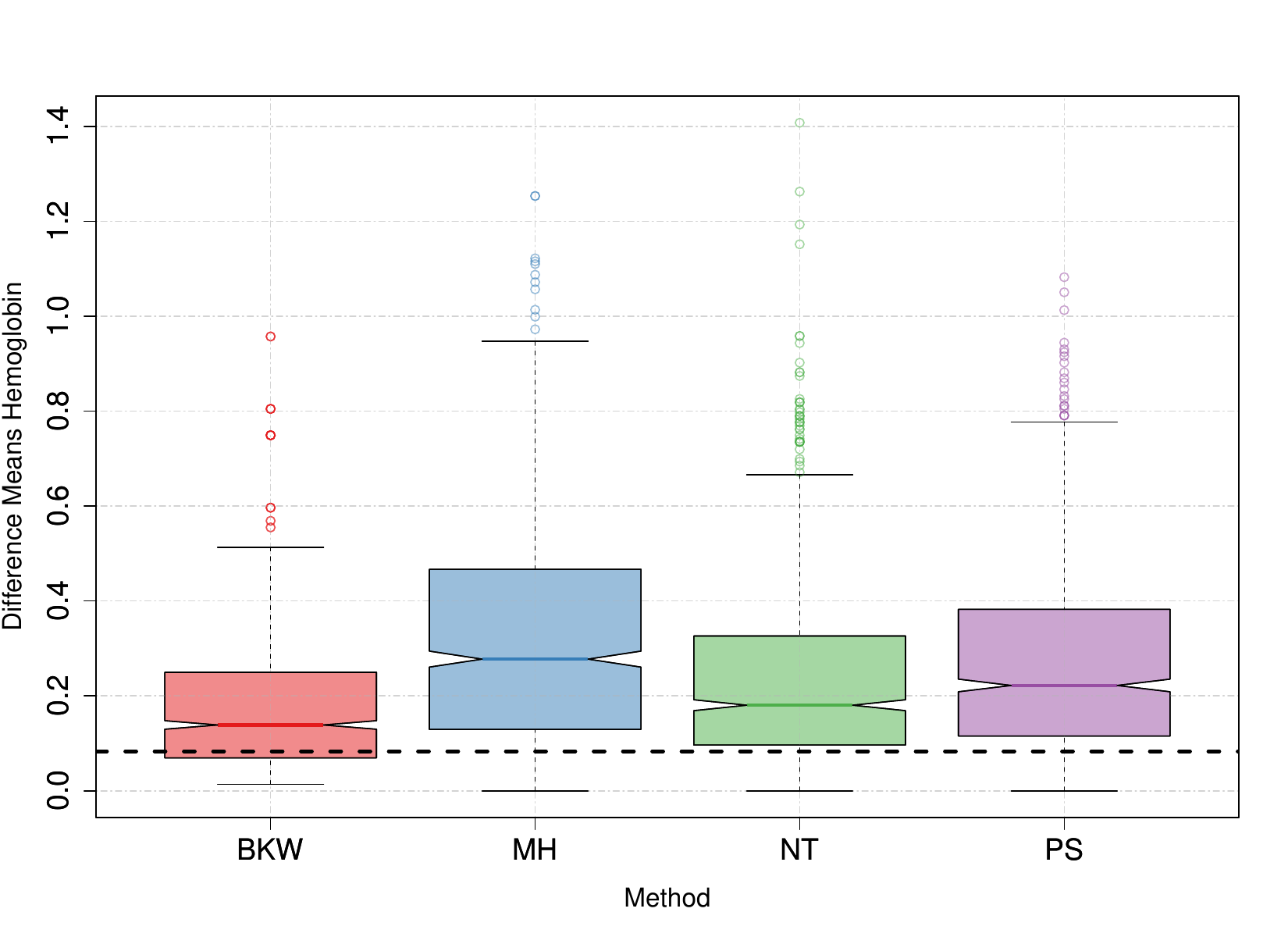}
         \caption{Absolute difference in means}
         \label{fig:pembro_hemoglobin_mean}
     \end{subfigure}
     ~
     \begin{subfigure}[b]{0.48\textwidth}
         \includegraphics[width=\textwidth]{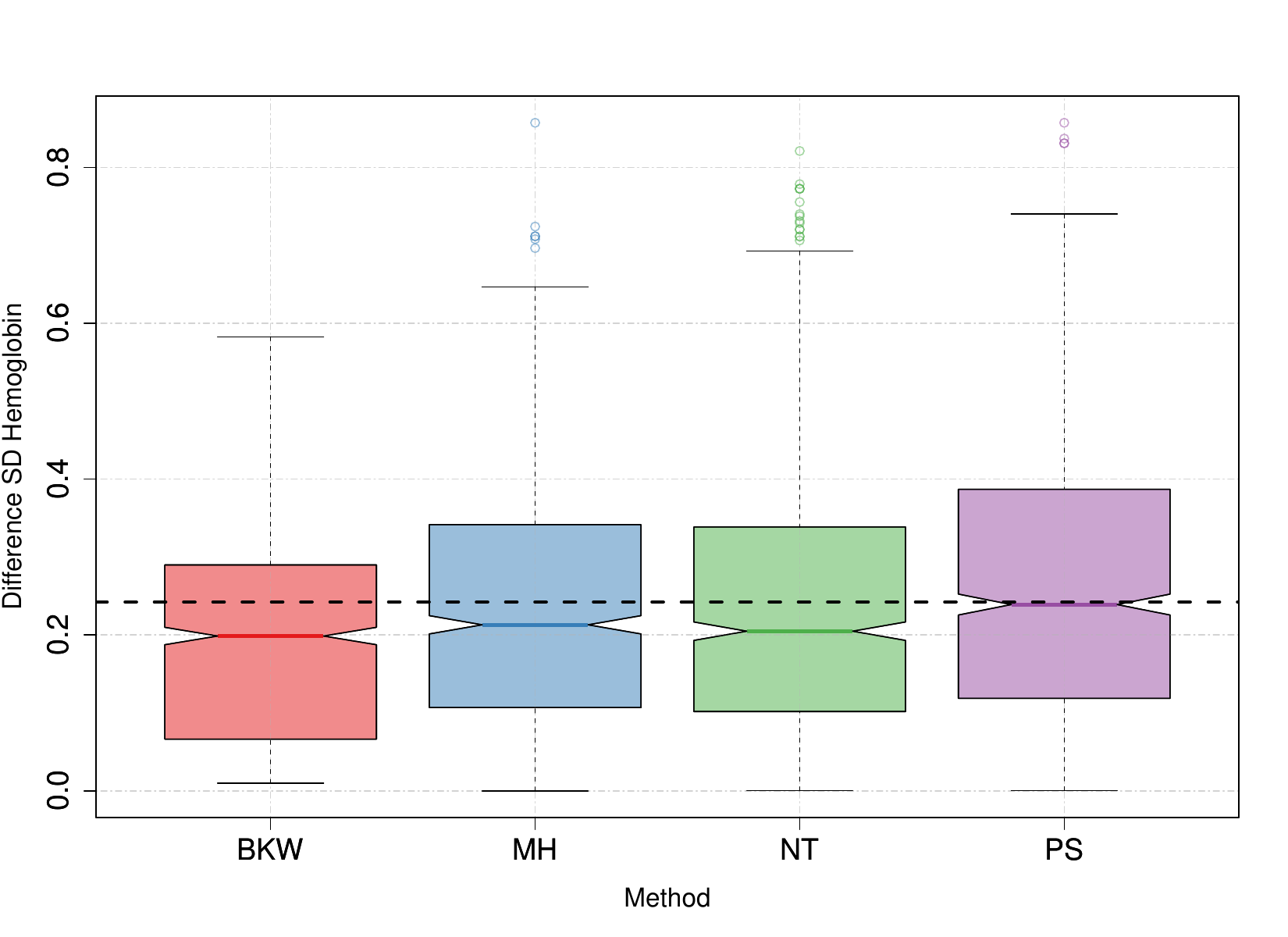}
         \caption{Absolute difference in standard deviations}
         \label{fig:pembro_hemoglobin_sd}
     \end{subfigure}
        \caption{Boxplots of the absolute differences in group means and standard deviations for hemoglobin in the pembrolizumab trial. The dashed line shows the actual absolute differences.}
        \label{fig:pembro_hemoglobin}
\end{figure}

\begin{figure}[h]
\centering
    \includegraphics[width=0.8\textwidth]{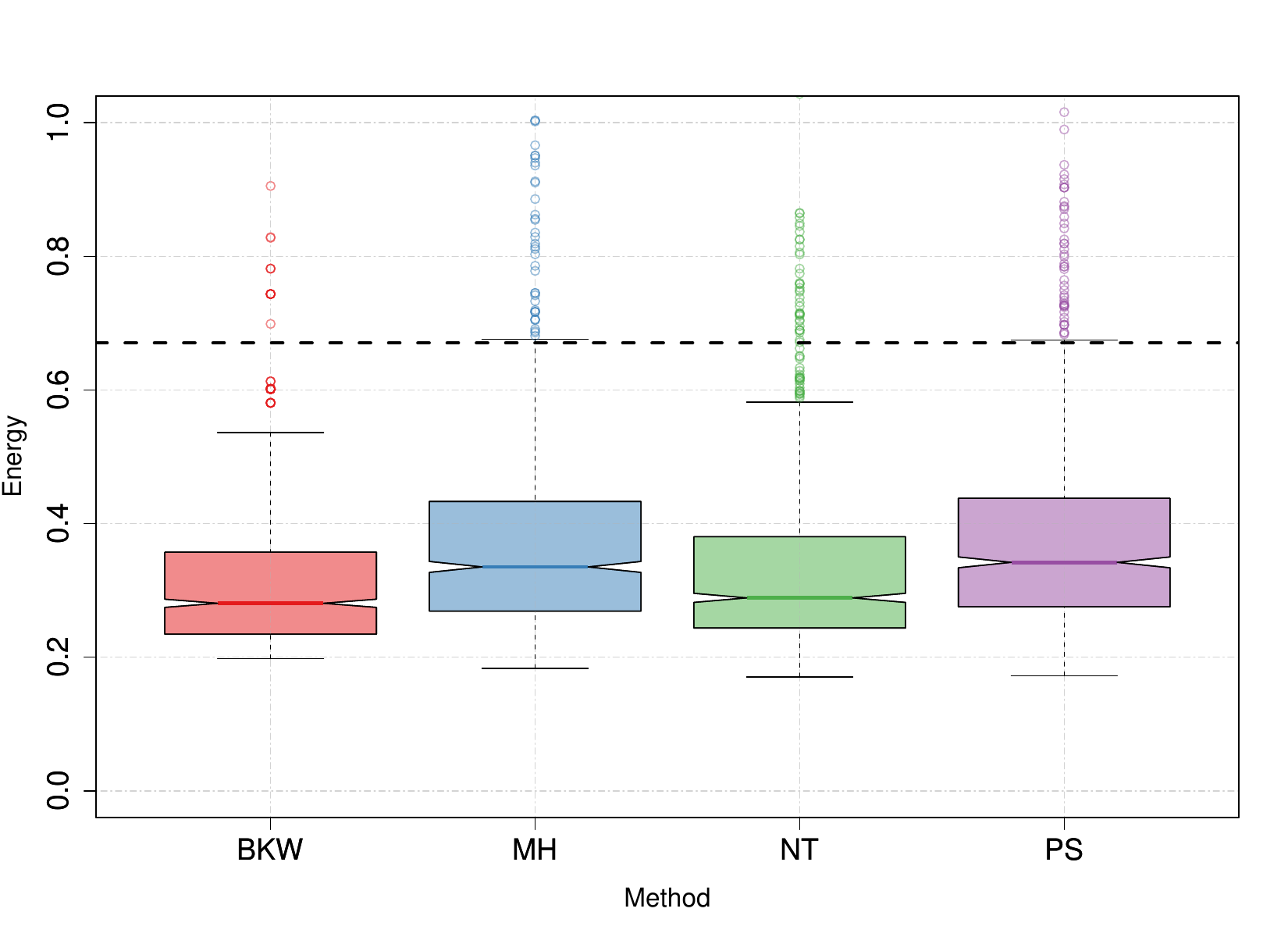}
    \caption{Boxplots of the energy distance values for the pembrolizumab trial. The dashed line shows the energy distance value observed in the trial.}
 \label{fig:pembro_energy}
\end{figure}

\begin{figure}[h]
\centering
    \includegraphics[width=0.8\textwidth]{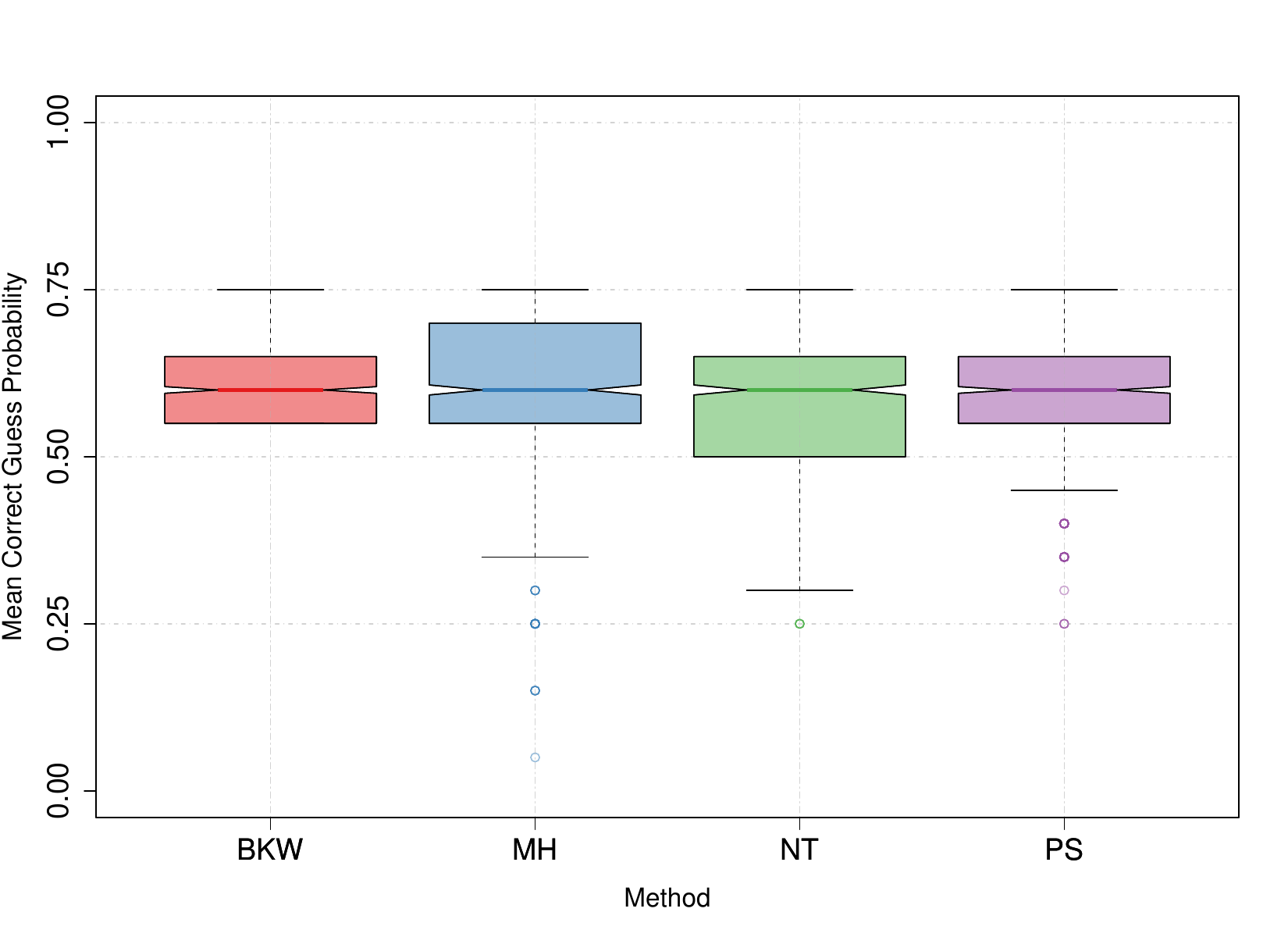}
    \caption{Boxplots of the mean correct guess probability values for the pembrolizumab trial.}
 \label{fig:pembro_CG}
\end{figure}

\reva{Figures~\ref{fig:pembro_group_balance} to \ref{fig:pembro_CG} show the results of the simulations. Specifically, Figures \ref{fig:pembro_group_balance}, \ref{fig:pembro_energy}, and \ref{fig:pembro_CG} show the distribution of the absolute difference between group sizes, the energy distance, and the mean CG probability, respectively, obtained by the methods. Figures \ref{fig:pembro_age}, \ref{fig:pembro_PD}, and \ref{fig:pembro_hemoglobin} show the distribution of the difference in the mean and standard deviations of the groups for age, modified proportion score of PD-L1, and hemoglobin concentration, respectively. Figures \ref{fig:pembro_age} to \ref{fig:pembro_energy} have a horizontal dashed line placed at the corresponding metric values observed in the clinical trial.}

Regarding the absolute difference between the group sizes, Figure~\ref{fig:pembro_group_balance} shows that the median absolute difference is two for the PS, NT, and MH methods. In contrast, the group absolute differences obtained by the BMW method are all equal to zero. This is because, in contrast to the other methods, the BMW method enforces group size balance.

Figure \ref{fig:pembro_age_mean} shows that all minimization methods improve the actual group assignment for the absolute difference between group means of age. This is because at least 75\% of the group mean differences for all methods are below the dashed line. The BKW method has a lower median and third quartile of group mean differences than the rest.

Figure~\ref{fig:pembro_age_sd} shows that all methods are better than the actual group assignment to balance the standard deviations of age, because at least 75\% of their absolute differences in the group standard deviations are below that obtained from the actual assignment. The NT method is better than the others because both its median and third quartile of group mean differences are smaller than those of the other methods.

Regarding the group means for PD-L1, Figure~\ref{fig:pembro_PD_mean} shows that all minimization methods improve the actual assignment of groups, since 75\% of their absolute group mean differences are below the observed in the clinical trial. In this case, the NT method is the most successful because it has a lower median absolute group difference than the others. Regarding the group standard deviations for PD-L1, Figure~\ref{fig:pembro_PD_sd} shows that at least 50\% of the group absolute differences obtained from the methods are smaller than the actual group difference in the clinical trial. The NT method is again better than the rest because its median absolute difference in group standard deviation is smaller than the others.

Regarding hemoglobin concentration, Figure~\ref{fig:pembro_hemoglobin_mean} shows that, to a large extent, all methods have greater absolute group mean differences than the one observed in the clinical trial. This is because either the first or second quartile of each method is above the dashed line in the figure. For the absolute group difference of the standard deviations, Figure~\ref{fig:pembro_hemoglobin_sd} shows that about 50\% of the group differences for all methods are below the actual group difference observed. In any case, the BKW method performs slightly better than the other methods for this covariate, as its median group differences are smaller than the others.

Figure~\ref{fig:pembro_energy} shows that all methods are better at balancing the joint distribution of the covariates between the two groups. This is because at least 75\% of the energy distance values for all methods are smaller than \reva{0.671, which was} observed from the actual group assignment. The BKW and NT methods are the best in terms of the energy distance, since they have a lower median energy distance value than the rest. However, the BKW method has slightly lower median and third quartile values than the NT method.

Regarding randomness in assigning subjects to groups, Figure~\ref{fig:pembro_CG} shows that the median values of the mean CG probability are the same for all methods. However, the NT method tends to have a smaller mean CG probability than the rest because its first and third quartiles are smaller than those of the others. In any case, the first quartile of the mean CG probability is at least 1/2 for all methods.

\subsection{The infant spasms clinical trial}
\label{sec:spasms_trial}

\cite{Chiron1997RandomizedSclerosis} studied the effect of vigabatrin on spasms due to tuberous sclerosis in infants. To this end, they conducted a clinical trial with 22 infants whose prognostic factors included age, sex, and duration and frequency of spasms. As an alternative treatment, \cite{Chiron1997RandomizedSclerosis} used hydrocortisone which is a standard steroid. In the trial, the infants were enrolled into two equally sized groups. One group was treated with vigabatrin while the other was treated with hydrocortisone. To assess the efficacy of the treatments, the authors used the time to disappearance of the spasms, the tolerability to the treatment, the evolution of the development quotient, among other responses.

\begin{figure}[h]
\centering
    \includegraphics[width=0.8\textwidth]{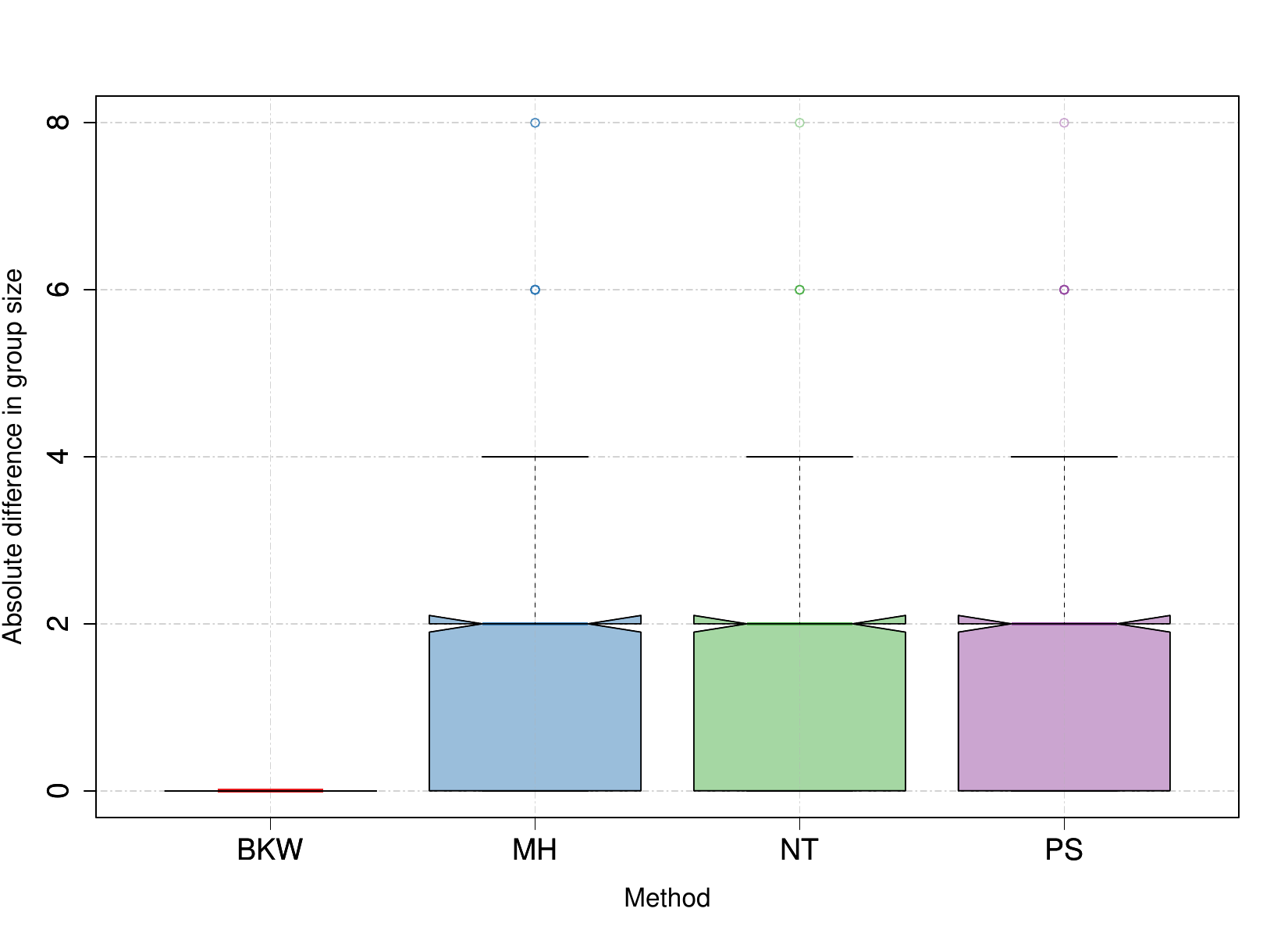}
    \caption{Boxplots of the absolute difference in group sizes obtained for the infant spasms trial.}
 \label{fig:infant_group_balance}
\end{figure}

\begin{figure}[h]
     \centering
     \begin{subfigure}[b]{0.48\textwidth}
         \includegraphics[width=\textwidth]{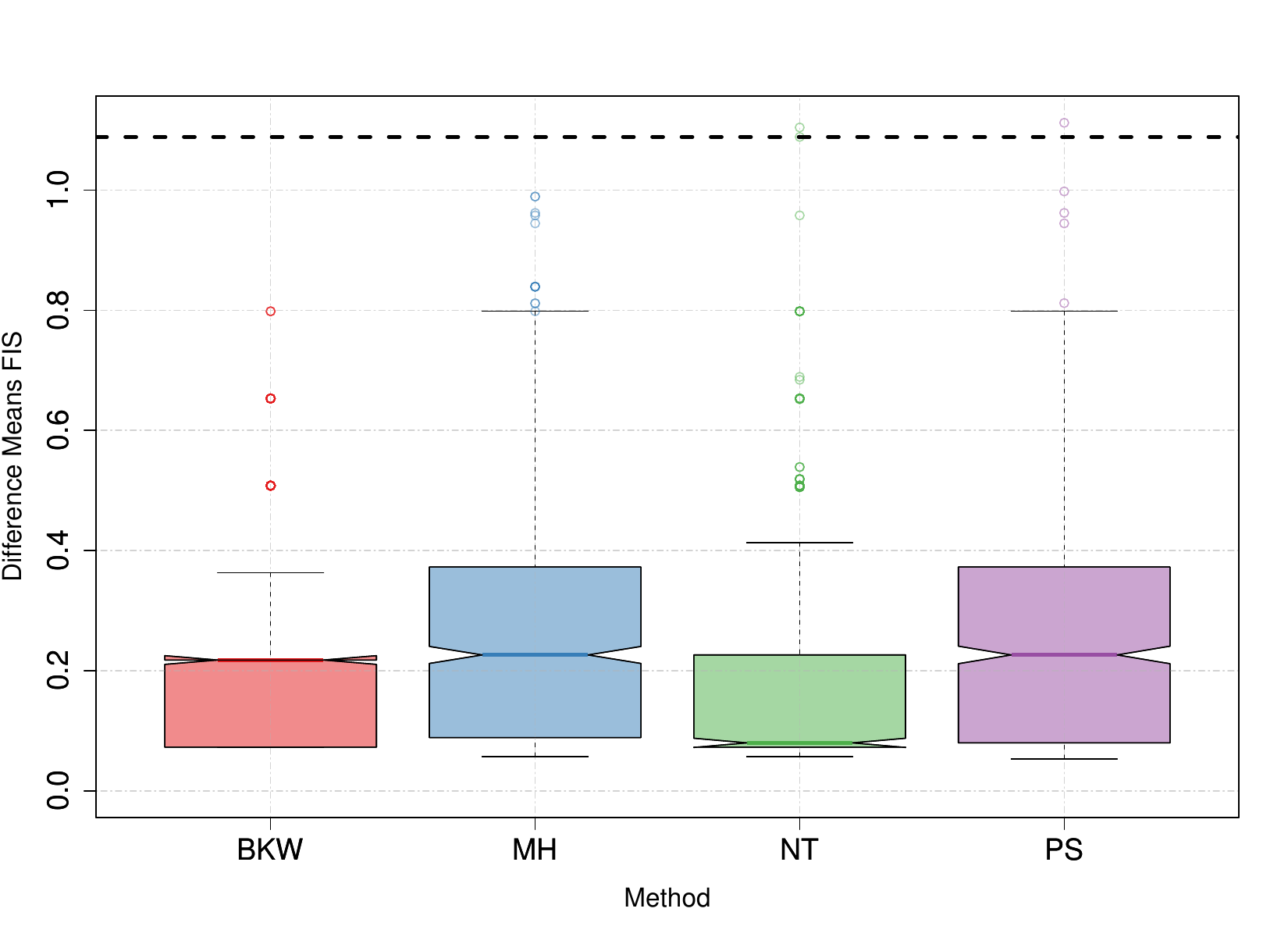}
         \caption{Absolute difference in means}
         \label{fig:spasms_fis_mean}
     \end{subfigure}
     ~
     \begin{subfigure}[b]{0.48\textwidth}
         \includegraphics[width=\textwidth]{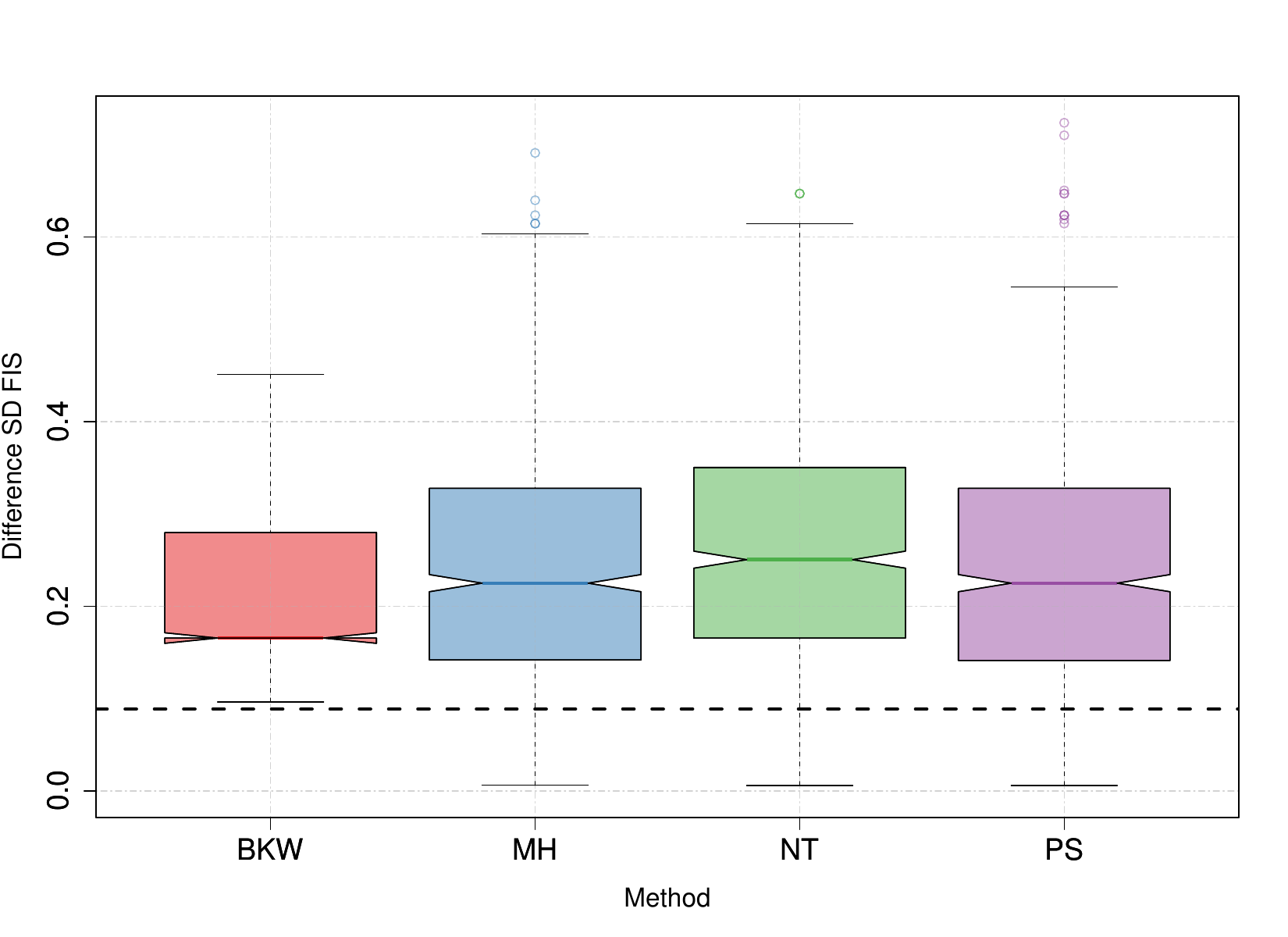}
         \caption{Absolute difference in standard deviations}
         \label{fig:spasms_fis_sd}
     \end{subfigure}
        \caption{Boxplots of the absolute differences in group means and standard deviations for FIS in the infant spasms trial. The dashed line shows the actual absolute differences.} \label{fig:spasms_fis}
\end{figure}

\begin{figure}[h]
     \centering
     \begin{subfigure}[b]{0.48\textwidth}
         \includegraphics[width=\textwidth]{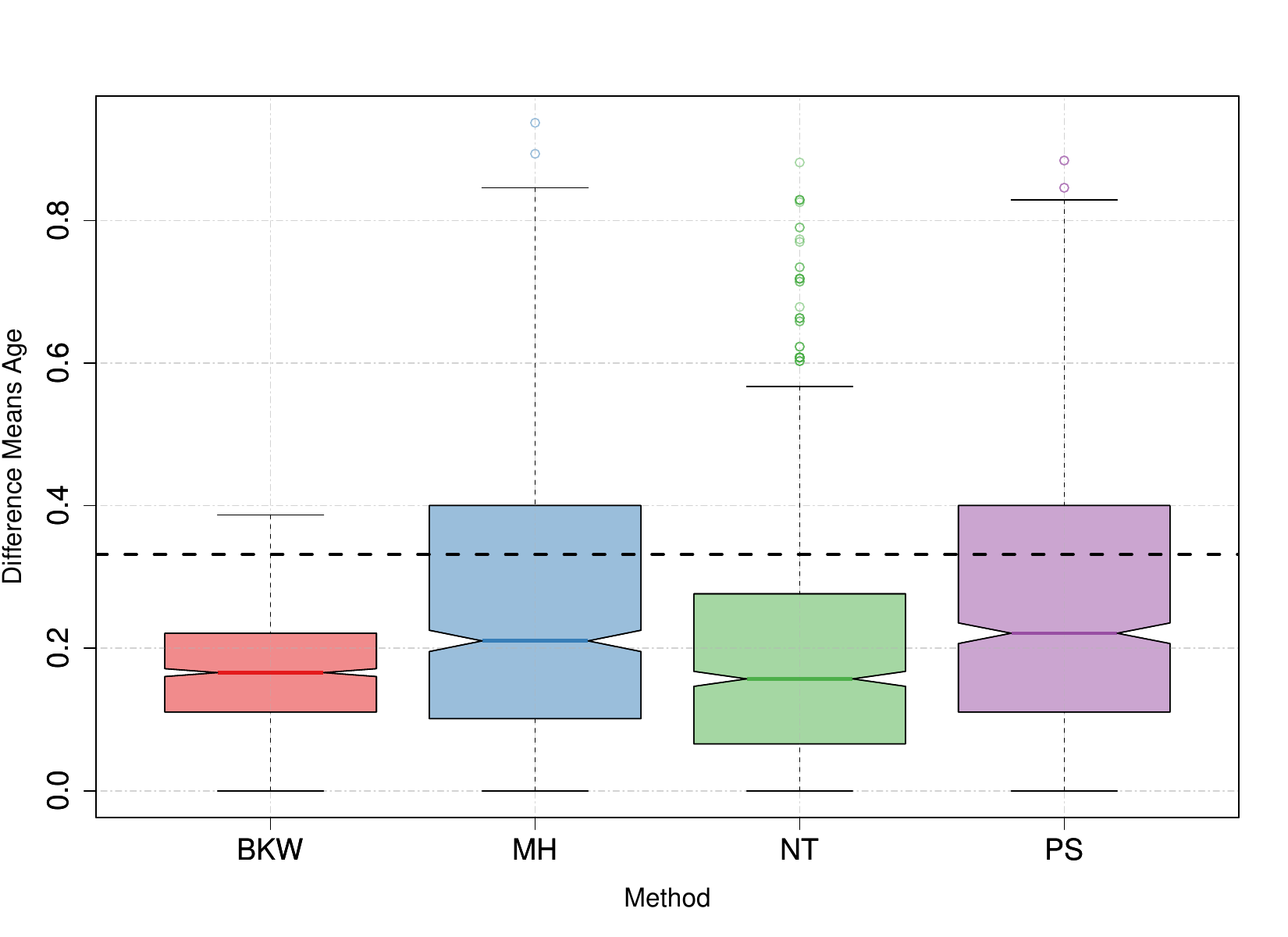}
         \caption{Absolute difference in means}
         \label{fig:spasms_age_mean}
     \end{subfigure}
     ~
     \begin{subfigure}[b]{0.48\textwidth}
         \includegraphics[width=\textwidth]{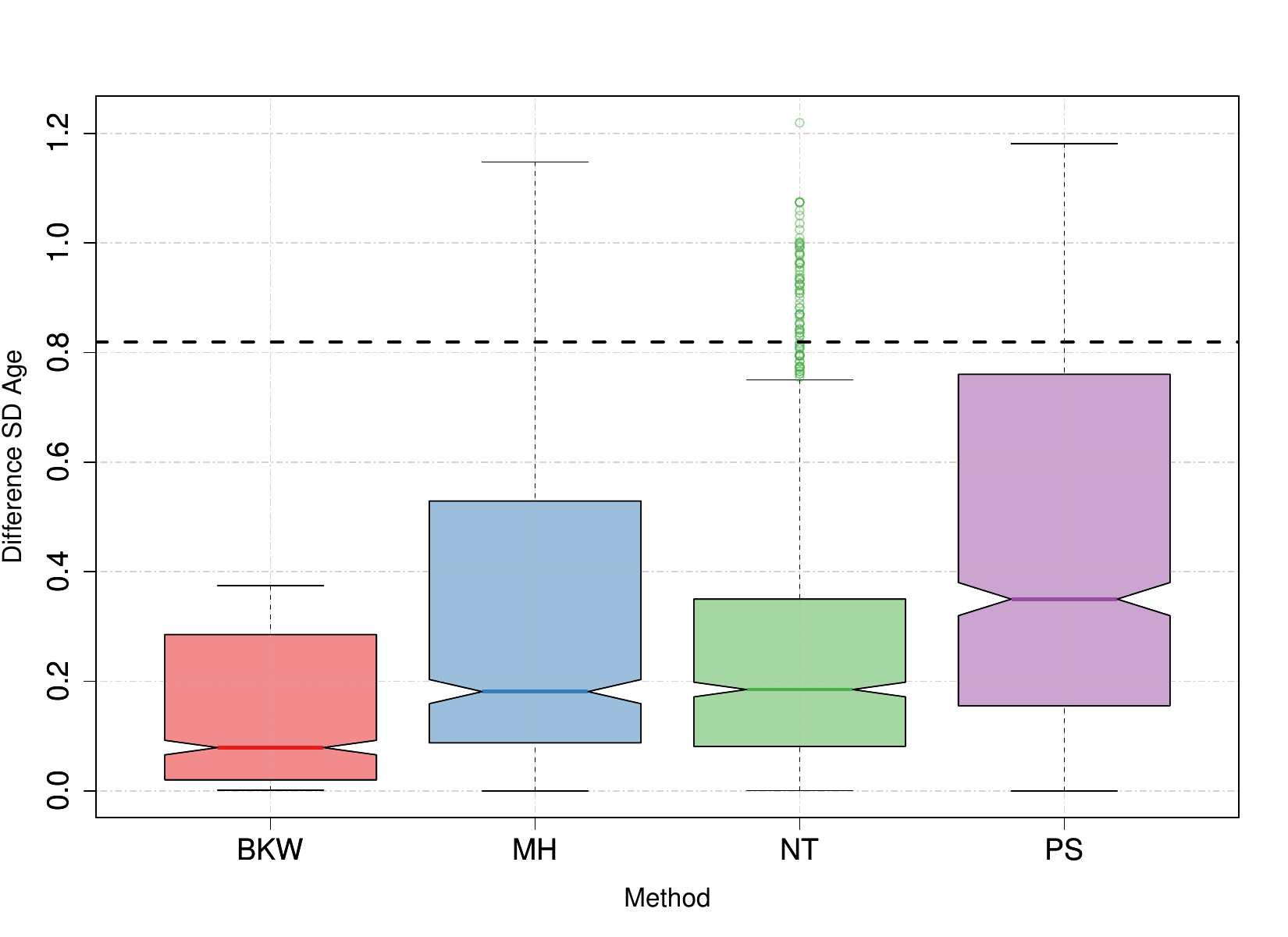}
         \caption{Absolute difference in standard deviations}
         \label{fig:spasms_age_sd}
     \end{subfigure}
        \caption{Boxplots of the absolute differences in group means and standard deviations for age in the infant spasms trial. The dashed line shows the actual absolute differences.}
        \label{fig:spasms_age}
\end{figure}

\begin{figure}[h]
\centering
    \includegraphics[width=0.8\textwidth]{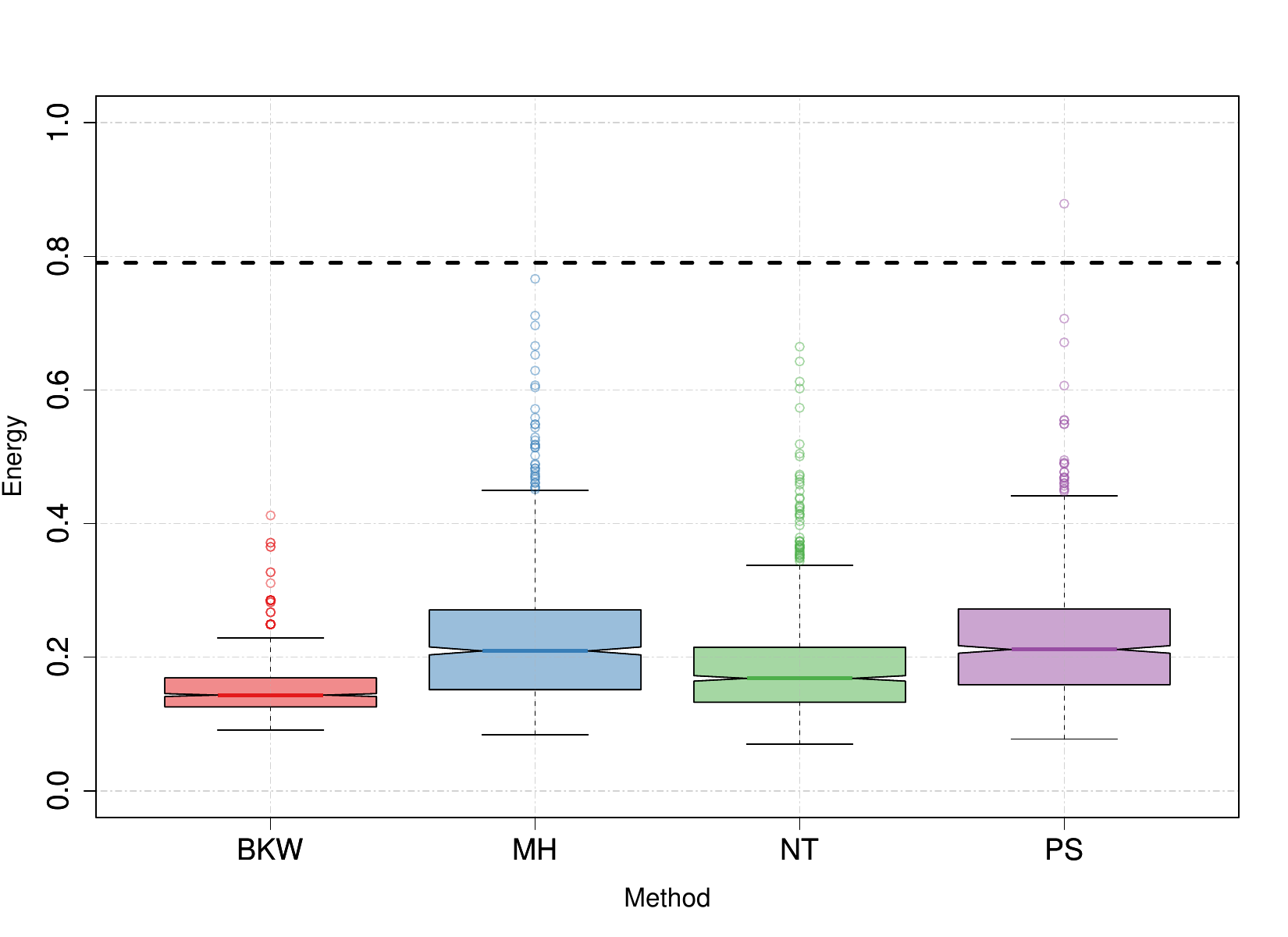}
    \caption{Boxplots of the energy distance values for the infant spasms trial. The dashed line shows the energy distance value observed in the trial.}
\label{fig:spasms_energy}
\end{figure}

\begin{figure}[h]
\centering
    \includegraphics[width=0.8\textwidth]{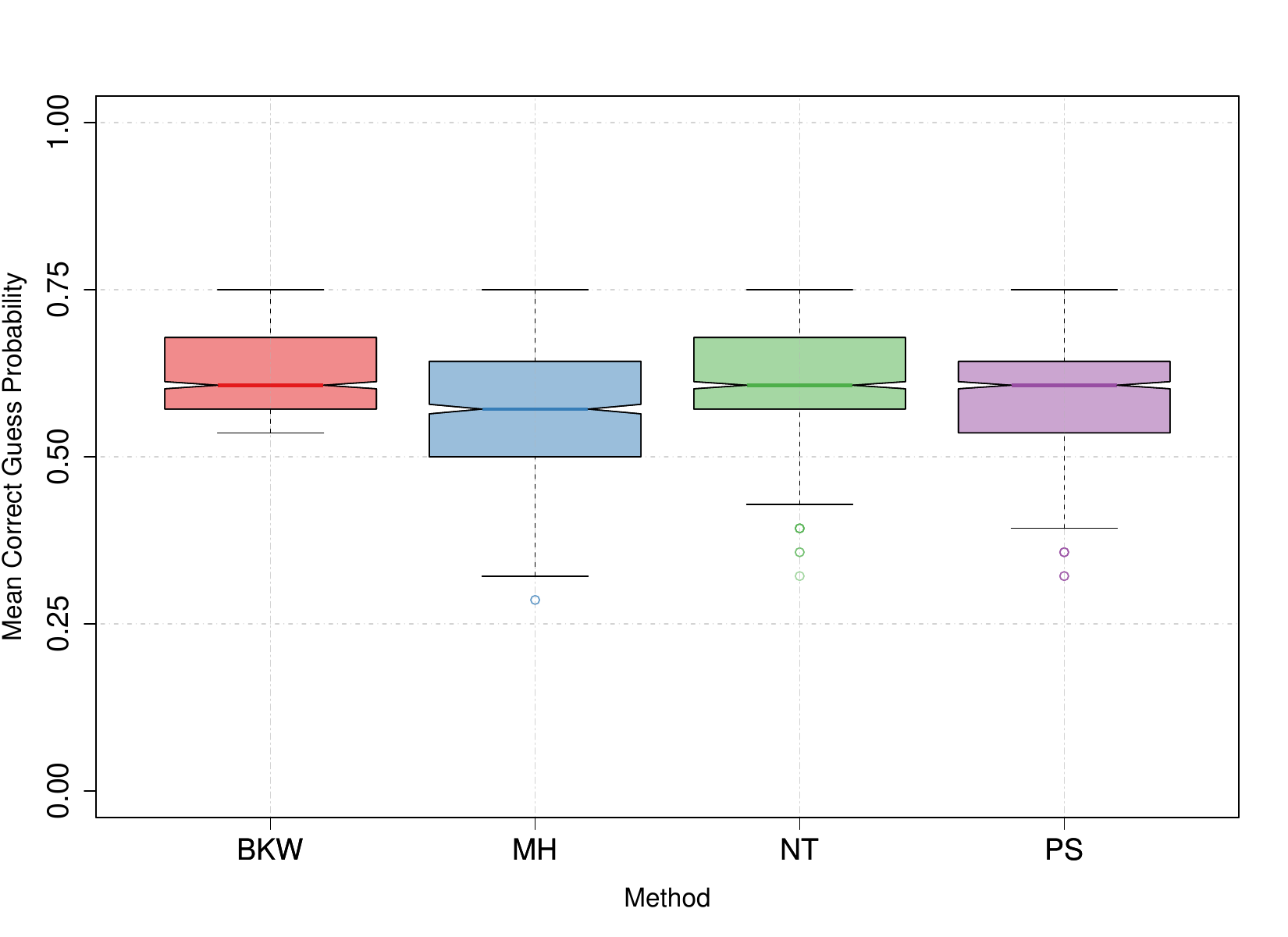}
    \caption{Boxplots of the mean correct guess probability values for the infant spasms trial.}
 \label{fig:spasms_CG}
\end{figure}

We compare the actual group assignment of \cite{Chiron1997RandomizedSclerosis} with minimization methods to balance the continuous covariates: frequency of infantile spasms (FIS) and age. Similarly to the pembrolizumab trial, we standardized these covariates to have a mean of zero and a standard deviation of one, and executed each method 1,000 times. \reva{Figures~\ref{fig:infant_group_balance} to \ref{fig:spasms_CG} visualize the simulation results for the infant spasms trial. Figures~\ref{fig:infant_group_balance}, \ref{fig:spasms_energy}, and \ref{fig:spasms_CG} are similar to Figures~\ref{fig:pembro_group_balance}, \ref{fig:pembro_energy}, and \ref{fig:pembro_CG} for the pembrolizumab trial. Figures~\ref{fig:spasms_fis} and \ref{fig:spasms_age} show the distribution of the difference in the mean and standard deviations of the groups for the FIS and age, respectively. Figures concerning the balance in covariate distributions have a horizontal dashed line placed at the observed metric values in the trial.} 

\reva{Regarding the balance in the group sizes, Figure~\ref{fig:infant_group_balance} shows that the BKW method is better than the others because all its group size differences are zero. In contrast, the other methods have a median absolute difference between group sizes equal to two.} 

Figure~\ref{fig:spasms_fis_mean} shows that all minimization methods improve the actual group assignment in terms of the mean of the FIS. This is because virtually all their absolute differences are smaller than the actual one indicated by the dashed line in the figure. The NT method has the smallest median absolute group difference. Regarding the absolute difference in the standard deviations for the FIS, Figure~\ref{fig:spasms_fis_sd} shows that the BKW method has smaller first and second quartiles than those of the other methods. However, none of the methods outperforms the actual group assignment in this case. This is because the first quartiles of the absolute group differences of all methods are greater than the difference observed in the clinical trial.  

Figure~\ref{fig:spasms_age} shows that the BWK and NT methods are generally better than the others in terms of balancing age, since at least 75\% of their group differences are below the differences observed in the trial. For the absolute difference in group means, the NT method has the lowest median, but the BWK has the smallest third quartile and dispersion; see Figure~\ref{fig:spasms_age_mean}. For the differences between group standard deviations, the BKW method has a median and third quartile that are smaller than those of the other methods; see Figure~\ref{fig:spasms_age_sd}.

Figure~\ref{fig:spasms_energy} shows that all methods are better than the actual clinical trial in terms of the energy distance. This is because the energy distance values of all the methods are smaller than the observed energy distance value\reva{, which was 0.79}. The BKW method is the best in this case, as its first and third quartiles are smaller than those of the other methods.

Figure~\ref{fig:spasms_CG} shows that the MH method has smaller mean CG probability values than the other methods. This is because all its quartiles are smaller than those of the others. However, none of the methods succeed in minimizing the mean CG probability, since more than 75\% of their values are higher than or equal to 1/2.

\subsection{Discussion} \label{subsec:discussion}

\reva{In Sections~\ref{sec:pembro_trial} and \ref{sec:spasms_trial}, we provided a detailed comparison of the minimization methods in terms of the balance in the group sizes and the marginal and joint distributions of the covariates, as well as the randomness in the assignment of subjects to groups. We presented the analysis separately for each evaluation metric. To help users select a minimization method, we constructed a radar plot \citep{Chambers1983} to visualize the overall performance of the methods in the trials. More specifically, the radar plot shows the average values for the absolute difference in group sizes, energy distance, and mean CG probability obtained by the methods. Each corner of the plot corresponds to the highest average value of a metric among all methods. Since lower average metric values are preferred, the points of a method should be close to the center of the plot.}

\begin{figure}[h]
     \centering
     \begin{subfigure}[b]{0.48\textwidth}
         \includegraphics[width=\textwidth]{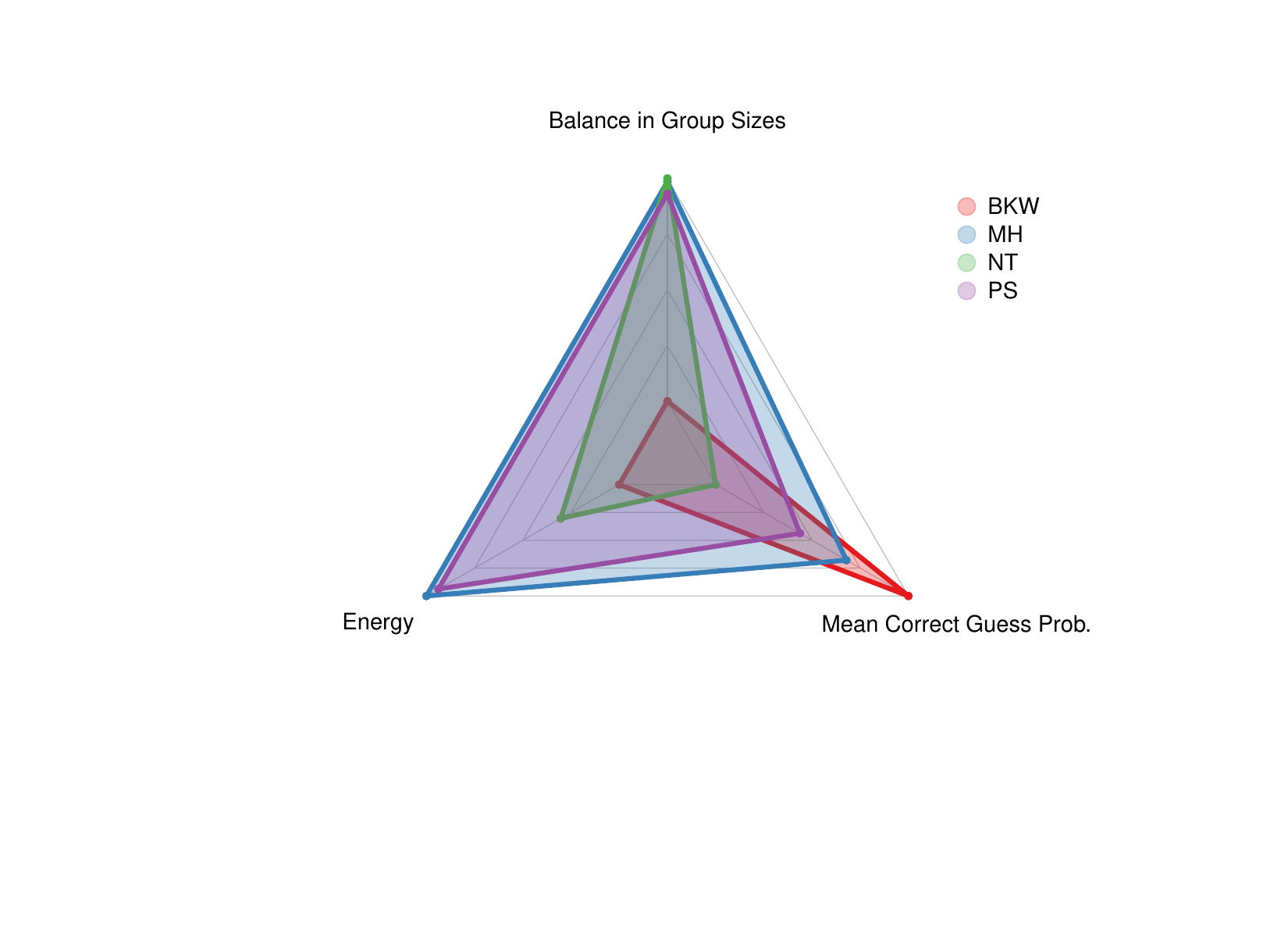}
         \caption{Pembrolizumab trial}
         \label{fig:pembro_Radar}
     \end{subfigure}
     ~
     \begin{subfigure}[b]{0.48\textwidth}
         \includegraphics[width=\textwidth]{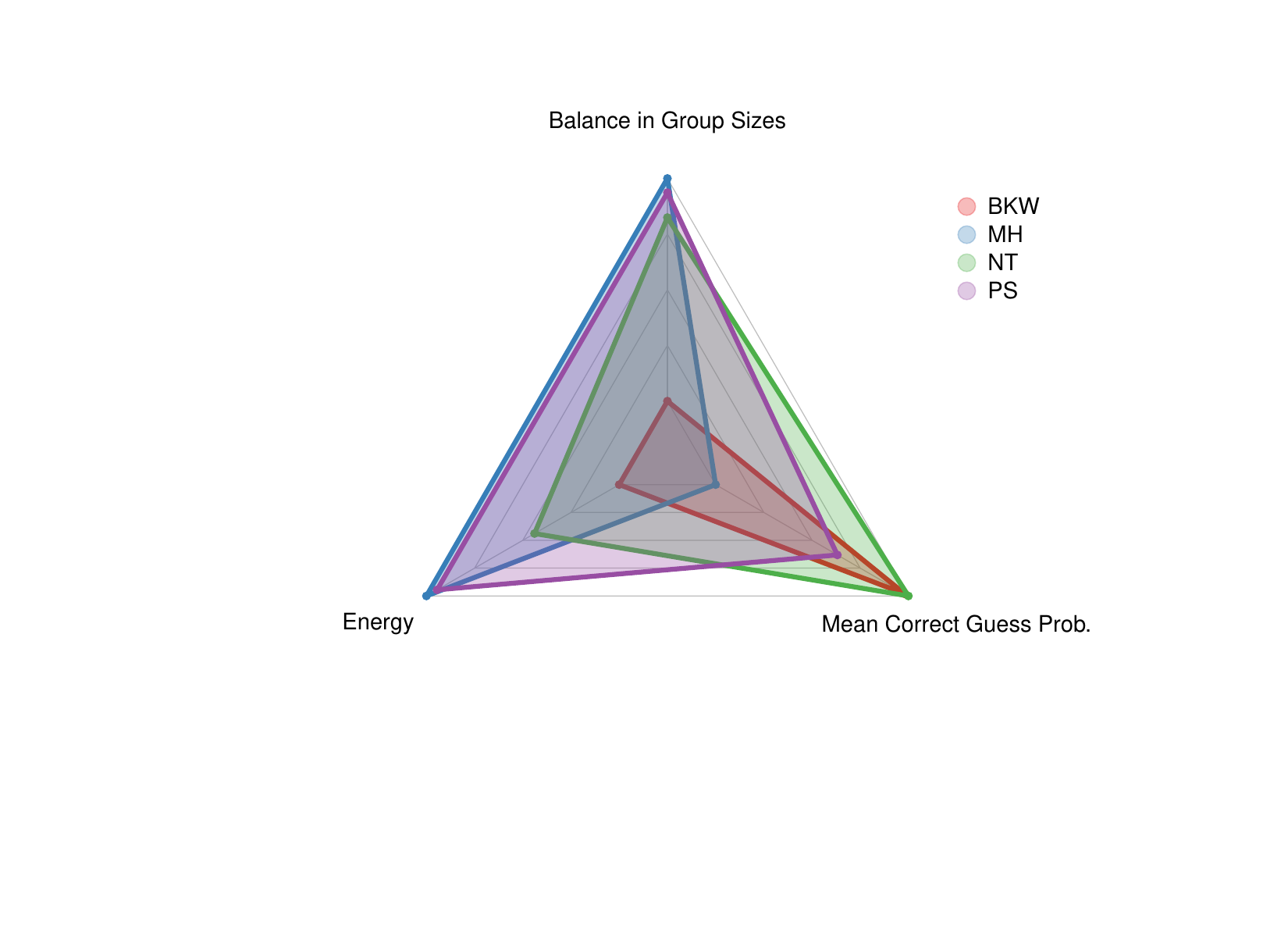}
         \caption{Infant spasms trial}
         \label{fig:spasms_Radar}
     \end{subfigure}
        \caption{Radar plot of the average values for the absolute difference in group sizes, energy distance, and the mean correct guess probability values for the pembrolizumab and infant spasms trial.}
        \label{fig:radar_plots}
\end{figure}

\reva{Figures~\ref{fig:pembro_Radar} and \ref{fig:spasms_Radar} show the radar plot for the pembrolizumab and infant spasms trial, respectively. On average, the BKW method outperforms the other methods in terms of the absolute difference in group sizes and covariate balance for both trials. This is because their average values for these metrics are closer to the center of the plot than those of the other methods. Therefore, we recommend the BKW method if the goal is to balance the distributions of the covariates. However, the BKW method has higher average values of the mean CG probability than the other methods in both trials. If the goal is to minimize the mean CG probability, we recommend the NT or MH method because they have the smallest average mean CG probability values for the pembrolizumab and infant spasms trial, respectively.} 

\reva{The conclusion of Figure~\ref{fig:radar_plots} on the BKW method agrees with the literature on minimization methods, since balance and allocation randomness are conflicting objectives \citep{Rosenberger2008HandlingTrials, Rosenberger2016}. In other words, a minimization method with a good balance in group sizes and covariate distributions tends to have a low degree of randomness in the assignment of subjects to groups. In contrast, a method with a high degree of randomness of allocation typically performs poorly in terms of balancing group sizes and covariate distributions. In any case, the summary of the covariate balance given by the energy distance allows us to easily see this phenomenon using the radar plots in Figure~\ref{fig:radar_plots}. We therefore recommend using the energy distance to evaluate the performance of minimization methods in terms of covariate balance.}

\reva{The method of \cite{Bertsimas2019Covariate-adaptiveTrials} has a tuning parameter $\rho$ that controls the trade-off between the balance in the group means and the group standard deviations of the covariates. In our study, we followed their recommendation and used $\rho = 6$, thus favoring the balance in the group standard deviations over that of the group means. Using additional experiments (not shown here), we studied the effect of the tuning parameter on the balance of marginal covariate distributions, the energy distance, and the correct guess probability for the two trials. We found that decreasing the value of $\rho$ generally results in a better balance between the group means of the covariates at the expense of an increase in the imbalance between the group standard deviations. The energy distance and the correct guess probability were not severely affected by the value of $\rho$. In any case, none of the values that we tested significantly improved all metrics simultaneously for both trials. We conclude that $\rho = 6$ works well for the clinical trials discussed here.}


\section{Concluding remarks}
\label{sec:conc}

We studied the mathematical programming method of \cite{Bertsimas2019Covariate-adaptiveTrials} to balance continuous covariates in small two-arm clinical trials. We showed that this method belongs to the class of minimization methods that are well known in the literature on clinical trials. We compared the method with the standard minimization methods of \cite{Pocock1975SequentialTrial}, \cite{Nishi2003AnGroups}, and \cite{Ma2013BalancingDensities}, in terms of how well they balance the marginal and joint distributions of the covariates between the groups, and the randomness in the allocation of future subjects. \reva{For our two case studies, we show that the method of \cite{Bertsimas2019Covariate-adaptiveTrials} outperforms the other methods in terms of the balance in the group sizes and covariate distributions. However, the method has a low degree of randomness in assigning subjects to groups, which may result in selection bias.} Remarkably, our study also shows that none of the methods is entirely successful for \reva{our case studies}, as they did not balance the marginal distribution of one of their variables. Therefore, better methods are needed to balance the covariate distributions in small trials.

\reva{The radar plots in Section~\ref{subsec:discussion} allow us to compare the minimization methods in terms of three important dimensions: balance in group sizes, covariate balance, and randomness in group assignment. However, another important dimension of a small clinical trial is its ethics \citep{GARATTINI2009792, nardini2014ethics}. For example, when studying a rare disease, a clinical trial in which the most promising treatment is assigned more frequently to subjects than the other treatment is more attractive in terms of ethics, compared to a completely randomized trial without selection bias. However, \cite{Rosenberger2008HandlingTrials} show that it is difficult to develop a clinical trial with a high degree of randomization that is in accordance with ethics, because these are conflicting objectives. To our knowledge, there are no mathematical programming approaches to obtain clinical trials with a clear trade-off between these dimensions. Therefore, more research is needed in this direction. To this end, we can explore multi-objective mathematical programming \citep{ALVES200799,ozpeynirci2010exact} to obtain a list of solutions (i.e., clinical trials) that are good in at least one of the objectives.}

\begin{center}
{\large\bf SUPPLEMENTARY MATERIAL}
\end{center}

\begin{description}

\item[Programs.zip:] R codes containing an implementation of the minimization methods and discrepancy functions in the article are available. The codes can be used to reproduce the numerical comparisons for the pembrolizumab and infant spasms trials. 

\end{description}

\bibliographystyle{apalike}
\bibliography{manuscript.bbl}
\end{document}